\documentclass[twocolumn,showpacs,preprintnumbers,amsmath,amssymb,prb]{revtex4}
\usepackage{t1enc,amstext,graphicx}

\begin{document}


\title{Influence of internal disorder on the 
superconducting state in the organic layered superconductor 
$\kappa$-(BEDT-TTF)$_2$Cu[N(CN)$_2$]Br}

\author{M. Pinteri\'{c}}
\email{mpinter@ifs.hr}
\affiliation{Institute of Physics, P. O. Box 304, HR-10001 Zagreb, Croatia\\
and Faculty of Civil Engineering, University of Maribor, SLO-2000 Maribor, Slovenia}

\author{S. Tomi\'{c}}
\author{M. Prester}
%
%
\author{\DJ. Drobac}
\affiliation{Institute of Physics, P. O. Box 304, HR-10001 Zagreb, Croatia}

\author{K. Maki}
\affiliation{Max-Planck Institute for the Physics of the Complex Systems,
D-01187 Dresden, Germany\\
and Department of Physics and Astronomy, University of Southern
California, Los Angeles, CA 90089-0484, USA}

\date{\today}

\begin{abstract}
We report high-sensitivity AC susceptibility measurements of the 
penetration depth in the Meissner state of the layered organic 
superconductor $\kappa$-(BEDT-TTF)$_2$Cu[N(CN)$_2$]Br.  We have 
studied nominally pure single crystals from the two different 
syntheses and employed controlled cooling procedures in order to 
minimize intrinsic remnant disorder at low temperatures 
associated with the glass transition, caused by ordering of
the ethylene moieties in BEDT-TTF molecule at $T_{\text{G}} = 75$~K.  
We find that the optimal cooling procedures (slow cooling of -0.2~K/h or 
annealing for 3 days in the region of $T_{\text{G}}$) needed to 
establish the ground state, depend critically on the sample 
origin indicating different relaxation times of terminal ethylene 
groups.  We show that, in the ground state, the behavior observed 
for nominally pure single crystals from both syntheses is 
consistent with unconventional $d$-wave order parameter.  The 
in-plane penetration depth $\lambda_{\text{in}}(T)$ is strongly 
linear, whereas the out-of-plane component 
$\lambda_{\text{out}}(T)$ varies as $T^2$.  In contrast, the 
behavior of single crystals with long relaxation times observed 
after slow (-0.2~K/h) cooling is as expected for a $d$-wave 
superconductor with impurities (\textit{i.e.} 
$\lambda_{\text{in}}(T) \propto \lambda_{\text{out}}(T) \propto T^2$) 
or might be also reasonably well described by the $s$-wave model.  
Our results might reconcile the contradictory findings previously
reported by different authors.
\end{abstract}


\maketitle

\section{Introduction}

Since the discovery of superconductivity in $\kappa$-BEDT-TTF based 
materials a decade ago\cite{Urayama:CHL:NCS,Kini:IOC:Br}, the question of pairing symmetry 
has remained as one of the most intriguing issues.  From the very beginning, 
these materials have attracted a lot of interest, not only because 
they achieved the highest superconducting (SC) transition temperatures 
($T_{\text{C}}$) among organic materials, but also because of their 
similarities to the high-temperature cuprate superconductors.  First, 
the $\kappa$-(BEDT-TTF)$_2 X$ (abbreviated as $\kappa$-(ET)$_2 X$) 
are strongly anisotropic, quasi-two-dimensional materials, with a very 
weak interplane coupling.  This feature is due to the crystalographic
structure, in which orthogonally aligned BEDT-TTF dimers form 2D 
conducting layers sandwiched between the polymerized anion ($X$) layers.  
Second, antiferromagnetic (AF) and SC phases occur next to one another, 
which suggests that electron correlations play significant role in 
the establishment of the ground state.  Indeed, the ground state of
$\kappa$-(BEDT-TTF)$_2$Cu[N(CN)$_2$]Cl material is an insulating AF 
phase with mildly canted spins\cite{Miyagawa:PRL:Cl,Pinteric:EJB:Cl}, 
while the ground state of $\kappa$-(BEDT-TTF)$_2$Cu[N(CN)$_2$]Br 
(abbreviated as $\kappa$-(ET)$_2$Br) and $\kappa$-(BEDT-TTF)$_2$Cu(NCS)$_2$ 
(abbreviated as $\kappa$-(ET)$_2$NCS)
is a SC phase.  In the former, the applied pressure supresses AF state
and stabilizes SC state\cite{Lefebvre:PRL:Cl}, whereas by
deuterization of $\kappa$-(ET)$_2$Br, the ground state is gradually pushed
from SC toward AF state\cite{Taniguchi:PRB:Br}.  The phase diagram is,
therefore, quite similar to that of the cuprates if doping is replaced by 
pressure or deuterization.  Third, the normal 
state has some properties that are distinct from conventional metals,
supporting the importance of the electron correlations.  In particular, the
Knight shift decreases significantly below about 50~K suggesting a 
suppression of density of states, that is the appearance of a pseudogap 
near the Fermi energy \cite{Wzietek:JP1:ET}.  A broad dip in the 
electronic density of states around Fermi energy was also observed by 
STM measurements below about the same temperature\cite{Arai:SYM:NCS}.  
Further, there is a peak in $1/T_1 T$ at about 50~K, where $T_1$ is the 
nuclear spin-lattice relaxation time, which suggests the presence of short 
range AF correlations \cite{Wzietek:JP1:ET}.  This peak disappears 
under pressure concomitantly with superconductivity.

The presence of significant electron correlations strongly favors the
possibility of an unconventional SC.  Results in favor of $d$-wave order 
parameter have been obtained by different experimental techniques.
$^{13}$C NMR measurements showed that the spin-lattice relaxation rate 
follows $T^{3}$ dependence at very low temperatures.  This result, 
together with the Knight shift experiment, provides evidence for spin singlet 
pairing with nodes in the gap\cite{Mayaffre:PRB:Br}.  The low temperature 
specific heat\cite{Nakazawa:PRB:Br}, as well as the thermal 
conductivity\cite{Belin:PRL:NCS}, also showed a power-law behavior 
$c_{\text{s}}(T) \propto T^2$ and $\kappa(T) \propto T$, respectively.
Further, magnetic penetration depth $\lambda(T)$, measured by microwave 
cavity perturbation\cite{Achkir:PRB:NCS}, muon-spin relaxation\cite{Le:PRL:ET}, 
tunnel diode oscillator\cite{Carrington:PRL:ET} and 
AC~susceptibility\cite{Kanoda:PRL:NCS,Pinteric:PRB:Br}, also displayed
the power law behavior, usually in the form of a $T$ and/or a $T^{2}$ 
behavior for $\lambda(T)$ at low temperatures.

Recent angle resolved measurements of SC gap structure using 
STM\cite{Arai:PRB:NCS} and thermal conductivity\cite{Izawa:PRL:NCS} 
clearly showed the fourfold symmetry in the angular variation, 
characteristic of the $d$-wave superconducting gap.  Both measurements
have revealed that nodes are directed along the directions rotated by
45 degrees relative to the in-plane crystal axes, indicating the
$d_{x^2-y^2}$-wave superconductivity.  Such a nodal structure indicates
that both Fermi surfaces (oval-shape quasi-two-dimensional hole cylinder 
band and an open quasi-one-dimensional band\cite{Xu:PRB:NCS}) should
participate in SC pairing in contrast with theoretical predictions
of superconductivity induced by AF spin fluctuations\cite{Louati:PRB:ET}.

In contrast, some other penetration depth 
studies\cite{Harshman:PRL:NCS,Lang:PRL:NCS,Dressel:PRB:ET}, as well as the
most recent specific heat measurements\cite{Elsinger:PRL:Br,Muller:PRB:NCS}, led to
results favoring conventional $s$-wave order parameter.  In particular,
a strong-coupling $s$-wave order parameter was claimed to be observed in
the latter experiments.

The question arises what is the source of the conflicting results
and how could this discrepancy be resolved.  As far as experimental
determination of $\lambda(T)$ in the mixed state is concerned, the complex
vortex dynamics might present a serious problem already at fields as low
as 70--300~Oe, as pointed out by Lee \textit{et al.}\cite{Lee:PRL:NCS}.  Further,
additional complications might be due to an order-disorder transition
that bears glassy features, taking place at $T_{\text{G}} \approx 75$~K
for $\kappa$-(ET)$_2$Br or at $T_{\text{G1}} \approx 70$~K and 
$T_{\text{G2}} \approx 53$~K for $\kappa$-(ET)$_2$NCS\cite{Akutsu:PRB:ET,Muller:PRB:ET}.
The transition region is situated between 65~K and 85~K, and between 45~K and 
75~K for $\kappa$-(ET)$_2$Br and $\kappa$-(ET)$_2$NCS systems, respectively.
The transition is ascribed to the gradual freezing down of the remaining 
motion of the ethylene groups of the BEDT-TTF molecules that are 
thermally activated at high temperatures between the two possible 
conformations.  That is, the relative orientation of the outer C-C 
bonds can be either eclipsed or staggered.  Upon lowering the temperature, 
the former and latter are adopted for $\kappa$-(ET)$_2$Br and 
$\kappa$-(ET)$_2$NCS, respectively.  X-ray diffraction measurements 
showed that at 125~K (CH$_2$)$_2$ groups are ordered in average in the 
whole bulk\cite{Geiser:PHC:ET}.  However, the passage through the region of glassy transition 
appears to play a crucial role regarding the level of residual intrinsic 
disorder at low temperatures.  This might be also due to the anomalous 
changes in thermal expansion behavior inside the same temperature 
region\cite{Kund:PHB:Br,Watanabe:JSJ:Br}.  Rapid cooling rates are reflected in 
smaller resistivity ratio between $T_{\text{G}}$ and $T_{\text{C}}$ 
and larger resistivity humps centered at about 
60~K\cite{Su:PRB:Br,Taniguchi:PRB:Br}.  The understanding of transport 
properties in the normal state is further complicated by the fact that 
the standard resistivity behavior that resembles a semiconducting state
above 100~K\cite{Buravov:JP1:Br}, whereas it becomes metallic below, 
is not reported for all syntheses\cite{Schweitzer}.  In the latter case,
samples display a metallic behavior in the whole temperature range between 
room temperature (denoted as RT) and $T_{\text{C}}$.  In addition, 
indications are given that some Cu(II) may replace regular Cu(I) during 
synthesis, affecting the resistivity behavior \cite{Montgomery:SYM:Br}.  
The correlation between different mean free path in samples of different 
syntheses and presence of Cu(II) ions was also 
suggested\cite{Mielke:PRB:Br}.  Further, sample dependence and relaxation 
effects were also observed in the magnetization measurement results for 
$\kappa$-(ET)$_2$Br\cite{Taniguchi:SYM:Br}.  Samples of one synthesis 
show two peaks in the magnetization \textit{vs.} field ($M-H$) curve, 
in contrast to crystals of another synthesis that show only one.  
Finally, the observed anomalous cooling rate dependence of $M-H$ curve 
was attributed to the change in the resistivity curves and remnant 
disorder in the sample.

In an attempt to reconcile the existing contradictions, and determine the
pairing symmetry of the genuine SC ground state, we have undertaken an 
investigation that covered a broad range of single crystals of various 
syntheses and in which the influence of thermal cycling and sample 
history was checked in carefully designed experiments.  We have performed
a high-resolution AC susceptibility measurements in the Meissner state
in the two field geometries, \textit{i.e.} when AC field was parallel and
perpendicular to the crystal planes.  A quantitative data analysis, we 
have elaborated with scrutiny to account for the demagnetization 
correction in the latter geometry, enabled us to get a 
full characterization of each sample under study clarifying in this way 
a previously suspected sample dependence.  Indeed, our experiments
unfold that cooling procedures in the region of the glass transition, 
necessary to establish the bulk SC ground state, critically depend on 
the sample origin.  In particular, our results demonstrate that the
low temperature state is critically determined by time scale of
experiment in the region of the glass transition, revealing in that way 
different relaxation times of terminal ethylene groups in samples from
two distinct syntheses studied in the most detail.  We show that in the
ground state, the behavior observed for nominally pure single crystals 
from both syntheses is consistent with unconventional $d$-wave order
parameter, that is the low-temperature in-plane and out-of-plane
superfluid density is proportional to $T$ and $T^2$, respectively.
However, the behavior of the superfluid density observed after standard
slow cooling for single crystals with long relaxation times deviates
from that found in the ground state.  It is as expected for a $d$-wave
superconductor with impurities or might be reasonably well described 
by the $s$-wave model.  These results may give a solution to a decade long 
mystery in regard to the symmetry pairing in $\kappa$-BEDT-TTF
superconductors.
A preliminary report was given in Ref. \onlinecite{Tomic:PHC:SC}.

\section{Experimental technique}

Measurements of the complex AC susceptibility ($\chi = \chi'+ i\chi''$) 
were performed using a commercial AC susceptometer 
(\mbox{CryoBIND}/Sistemprojekt, Zagreb).  The sensitivity of the system,
expressed in equivalent magnetic moment, was $\Delta m = 2 \cdot 10^{-9}$~emu 
in the broad temperature range between 1.5~K and $T_{C}$ \cite{cryo}.  
Measurements were performed with $H_{\text{AC}}$ = 14~mOe at $f$ = 231~Hz.  
Gold 0.07\% iron - copper thermocouple was used as a thermometer.  
The sample was placed in the upper one of the two identical secondary 
coils immersed in the liquid helium bath, positioned in such a way 
that applied AC field was either perpendicular or parallel to the 
conducting planes of the single crystals under study.  In the remainder
of the paper, the case of magnetic field aligned with the 
b~crystallographic axis, \textit{i.e.} perpendicular to the high 
conducting ac crystallographic planes, is denoted by \label{H-plane}
$H_{\text{AC}} \perp$ plane, while the case of magnetic field direction 
laying inside the ac~crystallographic plane is denoted by 
$H_{\text{AC}} \parallel$ plane.  In order to probe the sample in the 
Meissner state care was taken to reduce the amplitude of the AC field 
($H_{\text{AC}}$) until the component $\chi'(T)$ was independent of 
$H_{\text{AC}}$ ($H_{\text{AC}} <$ 42~mOe) and the $\chi''(T)$ component 
was negligible.  No frequency dependence (13~Hz $< f <$ 2~kHz) was 
observed for $H_{\text{AC}} <$ 1~Oe. In addition no influence of the 
Earth's field was observed:  runs performed with compensation of the Earth's 
field gave the same results.  This is in accordance with the fact that 
the reported values for lower critical magnetic field $H_{C1}(T)$, 
corrected for demagnetization, are far above the Earth's field 
$H_{\text{E}}$ for all temperatures below 8~K.  In this temperature 
range $H_{C1}(T) \ge 10$~Oe \cite{Hagel:PHC:Br}, while the value of 
the Earth's field determined in our laboratory is 
$H_{\text{E}} \approx 0.36$~Oe \cite{Drobac:JMM:tech}.

\begin{figure}
\centering\includegraphics[clip,scale=1.00]{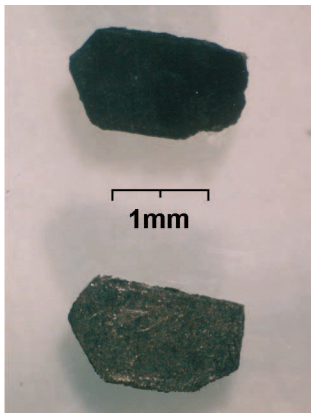}
\caption{The $\kappa$-(ET)$_2$Br (above) and its reference sample (below).}
\label{rs}
\end{figure}
The calibration of the system was performed with a piece of specially and 
carefully designed Niobium foil, which we will refer to, in the remainder of 
the paper, as the reference sample.  For each $\kappa$-(ET)$_2$Br single 
crystal, a special reference sample was created independently to ensure that 
its dimensions and the shape were as close as possible to the original (see 
Fig.~\ref{rs}).  We start with the conventional formula, which relates
the induced voltage on the detection coil $U$, the measured susceptibility 
$\chi'_{\text{m}}$, the demagnetization factor $D$, the volume $V$ 
and the internal susceptibility $\chi'$, that is $\chi'_{\text{m}}$ 
corrected for the demagnetization effect $1/(1+D\chi')$:
\begin{equation}
U \propto V \chi'_{\text{m}} = V \frac{\chi'}{1+D\chi'}.
\end{equation}
Here we have taken into account that measured samples are much smaller than
the detection coil.  If we use a conventional calibration assumption that 
the susceptibility for Niobium reference sample at low temperatures can be
taken to be -1, we can finally determine the internal value of the 
susceptibility according to formula
\begin{equation}
\chi' = - \frac{V_{\text{r}}}{V} \cdot \frac{\frac{1}{1-D_{\text{r}}}}{\frac{1}{1+D \chi'}} \cdot \frac{U}{U_{\text{r}}}
\label{cal4}
\end{equation}
where index r designates the reference sample.
When the AC~field was parallel to the crystal planes, the demagnetization 
effect could be neglected, meaning that the middle factor in 
Eq.~(\ref{cal4}) can be simply replaced by unity.  In order to correct 
the measured data for the demagnetization factor for the AC~field perpendicular
to the crystal planes, we have developed a method, which strongly 
reduces the experimental error in $\chi'$ to less than 1\% and allows 
us to get the reliable absolute value of the penetrated volume $1+\chi'$ 
for this geometry, as well.  The latter quantity directly determines the 
penetration depth and the superfluid density, as we show in Sections
\ref{lam} and \ref{rho}.  In the following, we argue and show that our 
claim is justified.

Our starting point is that the reference sample represents a perfect 
copy of the sample under study (Fig.~\ref{rs}), implying very close 
values of respective demagnetization factors $D$ and $D_{\text{r}}$.  
Since the $\kappa$-(ET)$_2$Br sample is almost completely 
diamagnetic in this geometry, we can use the approximation $\chi' \approx -1$.  
Therefore, demagnetization effects for both samples must be very close in 
value and the middle factor in Eq.~(\ref{cal4}) in the first 
order of approximation can be replaced by unity.  Further, in order to improve 
precision and eliminate a small remaining difference in demagnetization 
effect, the middle factor in the second order of approximation is calculated
by the following procedure.  Both $\kappa$-(ET)$_2$Br and the reference sample were 
taken to be fully superconducting disks (that is thin cylinders with aspect 
ratio of the length and the diameter of about 0.4) with the same face area and the 
same thickness as their originals.  We consider the disk approximation to be
more suitable for description of the real sample than the elipsoid one, used
by authors of Ref.~\onlinecite{Taniguchi:PRB:Br}.  We base this assertion 
on the fact that the former approximation better describes rather
sharp sample edges, which might give a substantial contribution to the
demagnetization factor.  The middle factor in Eq.~(\ref{cal4})
is then given by the calculated ratio of diamagnetic effects for these two
bodies, $\frac{1}{1-D_{\text{disk,r}}}/\frac{1}{1-D_{\text{disk}}}$.
The systematic error due to the approximation of the specific shape of
samples to the shape of the disk obviously cancels out by division.
Numerical data for the demagnetization factor for the disk were taken from 
the literature \cite{Crabtree:PRB:tech}.

In order to calculate the area and the thickness, dimensions of both $\kappa$-(ET)$_2$Br 
and reference sample were carefully measured with a high precision of 1\%.  
The precision was verified by the following procedure.  All significant 
Niobium reference sample dimensions were measured and the volume calculated.  
In addition, the same reference sample was weighted and the volume 
calculated using Niobium density, 
$\rho_{\text{Nb}} = 8.57$~g/cm$^3$.  The difference between two obtained 
values was always about 1\%.  Finally, the middle factor in Eq.~(\ref{cal4}) 
obtained in this way differs from unity only for a few percent.  This 
shows that the calculation procedure and the starting assumption are valid.

In the end, we estimate that our calibration procedure for
$H_{\text{AC}} \perp$~plane is accurate in $1 + \chi'$ to 33\%
for $\chi'$ close to -1, and to about 50\% for $-0.96 > \chi' > -0.5$.
As for the $H_{\text{AC}} \parallel$~plane, accuracy in $1 + \chi'$ is
estimated to be about 15\% for all measured low-temperature $\chi'$ values.

Nominally pure single crystals of $\kappa$-(ET)$_2$Br, originating from two 
different syntheses had different resistivity ratios 
$RR(75\textrm{K}/T_{\text{min}}$) for similar cooling rates 
employed\cite{Kanoda,Schweitzer}. 
$RR(75\textrm{K}/T_{\text{min}}) = \rho(75\textrm{K})/\rho_{\text{min}}$, 
where $\rho_{\text{min}}$ is the resistivity measured at temperature just 
above the SC transition.  Single crystals with $RR(75\textrm{K}/T_{\text{min}})$ 
$\approx 200$ and $\approx 50$ obtained by standard slow cooling 
rate have been labelled as~S1 and~S2, respectively.

Three different cooling procedures were used to cool samples from 
RT to 4.2~K.  Special care was taken in 
the temperature range $60~\textrm{K} < T < 100~\textrm{K}$, where 
relaxation processes appear\cite{Akutsu:PRB:ET,Muller:PRB:ET}.  For 
\textit{Quenched} (denoted as Q) state the sample was cooled down 
to liquid helium temperature in about one minute, which 
represents the average speed of about -300~K/min.\ over the whole 
temperature range.  For \textit{Relaxed} (denoted as R) state the
sample was first cooled to 100~K in about 10 minutes.  Between 
60~K and 100~K the cooling rate was carefully monitored to amount 
to about -0.2~K/min.  Below 60~K, the sample was finally cooled to 
4.2~K in a few minutes.  For the \textit{Annealed} state (denoted as A) 
the sample was cooled down to liquid nitrogen temperatures in 
about one hour.  Then it was kept between liquid nitrogen 
temperature and 100~K for three days.  Finally, it was cooled down
to liquid helium temperature in a few minutes.
Samples used in this study were rhombic platelets with face areas between 
0.51~and 2.15~mm$^{2}$ and thickness between 0.29~and 0.70~mm.

\section{Results and analysis}

\subsection{Complex susceptibility}

Our first important result concerns the influence of cooling rate 
on the components of the complex susceptibility in the SC state 
as a function of synthesis procedure.  We present the behavior 
obtained in two principal field geometries.  That is 
$H_{\text{AC}} \perp$ plane and $H_{\text{AC}} \parallel$ plane,
as defined in Section \ref{H-plane}.  On the basis of our 
previous measurements with the magnetic field aligned with the a~and 
c~crystallographic axes\cite{Pinteric:PRB:Br}, we know that the 
specific orientation within the ac~crystallographic plane does 
not influence the obtained result for the penetration depth since
the out-of-plane component is a dominant contribution.
Therefore, we always made sure that the field was aligned with 
the largest dimension of the platelet face in order to minimize 
the demagnetization factor.

\begin{figure}
\centering\includegraphics[clip,scale=0.50]{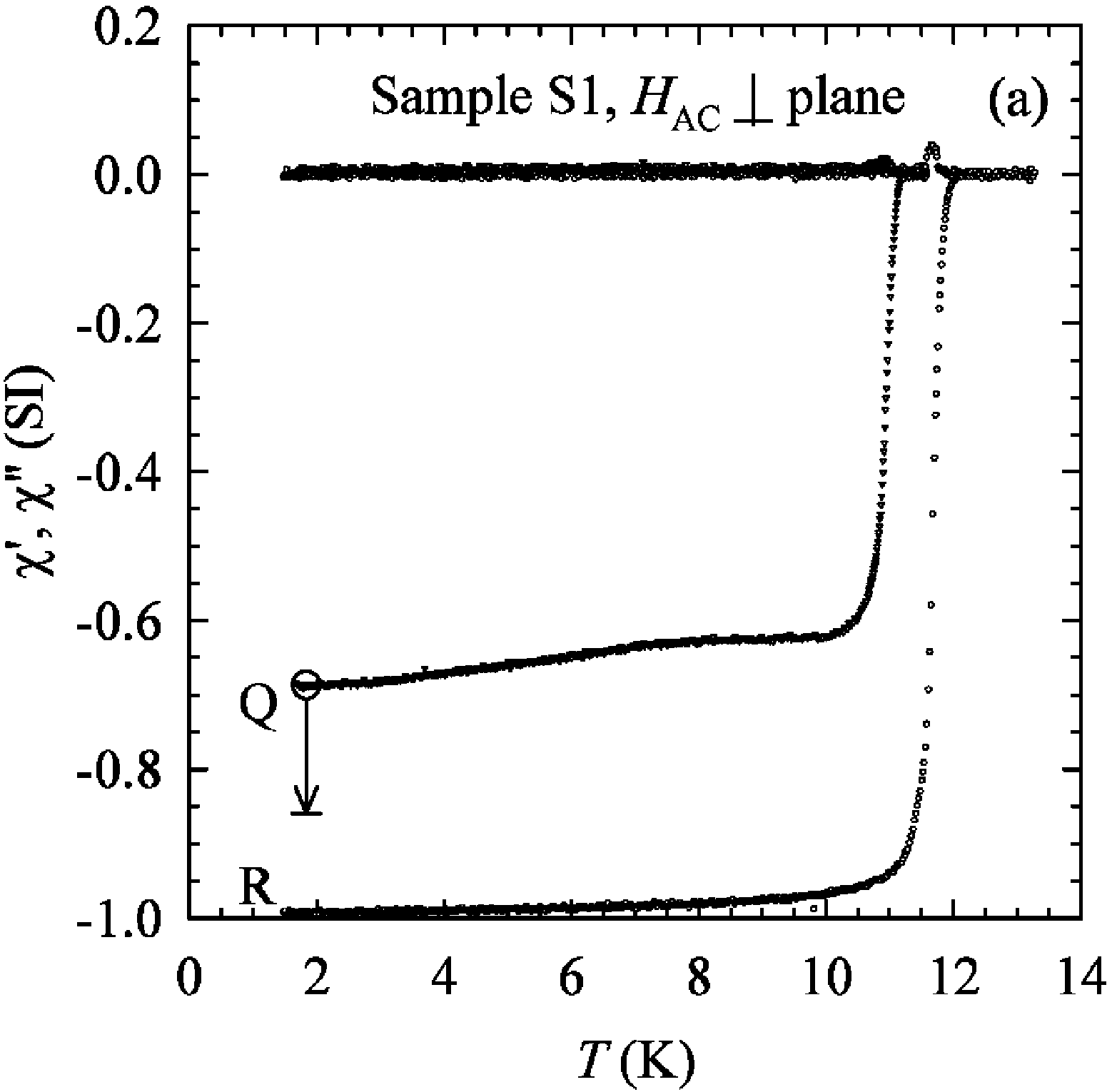}
\centering\includegraphics[clip,scale=0.50]{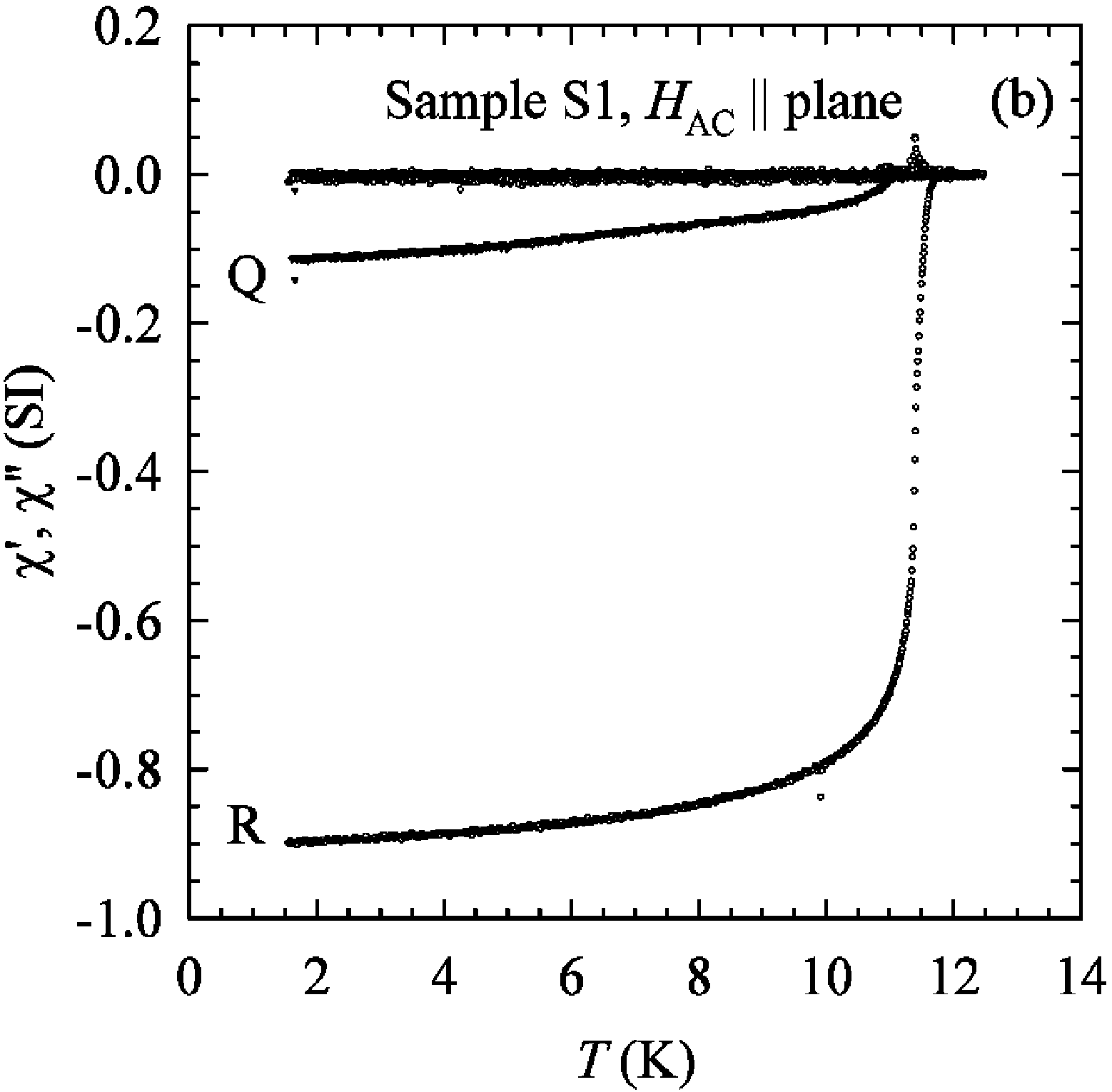}
\caption{Sample~S1: Real and imaginary part of susceptibility for Relaxed (R) 
and Quenched (Q) state in (a)~$H_{\text{AC}} \perp$ plane and
(b)~$H_{\text{AC}} \parallel$ plane geometry.  The arrow in (a) 
illustrate the upper limit of the systematic error.}
\label{cKan}
\end{figure}
Susceptibility data obtained for sample~S1 for two different cooling 
rates are shown in Fig.~\ref{cKan}.  We identify the R~state, as the 
\textit{ground state}.  In the ground state, superconductivity 
sets-in at 12.0~K and 11.5~K, for $H_{\text{AC}} \perp$ plane and 
$H_{\text{AC}} \parallel$ plane, respectively.  We have already reported 
difference in $T_{C}$, as well as anisotropy in the susceptibility for 
$H_{\text{AC}}$ perpendicular and parallel to the planes\cite{Pinteric:PRB:Br}.  
Here we report, to the best of our knowledge for the first time, the 
absolute susceptibility values in both field geometries.  For 
$H_{\text{AC}} \perp$ plane the sample response is almost completely 
diamagnetic, while for $H_{\text{AC}} \parallel$ plane the susceptibility 
is somewhat smaller in magnitude.  Next, our new result points to a huge 
effect of cooling rate on the susceptibility value and $T_{\text{C}}$.  
When sample was cooled faster, the absolute value of susceptibility 
was smaller, and $T_{\text{C}}$ lower.  In other words, the diamagnetic 
region shrinks in $\chi'$~\textit{vs.}~$T$ plot.  This huge effect is 
especially emphasized for $H_{\text{AC}} \parallel$ plane geometry,
where the absolute $\chi'$ value is almost an order of magnitude smaller 
for Q~state than for R~state.

\begin{figure}
\centering\includegraphics[clip,scale=0.50]{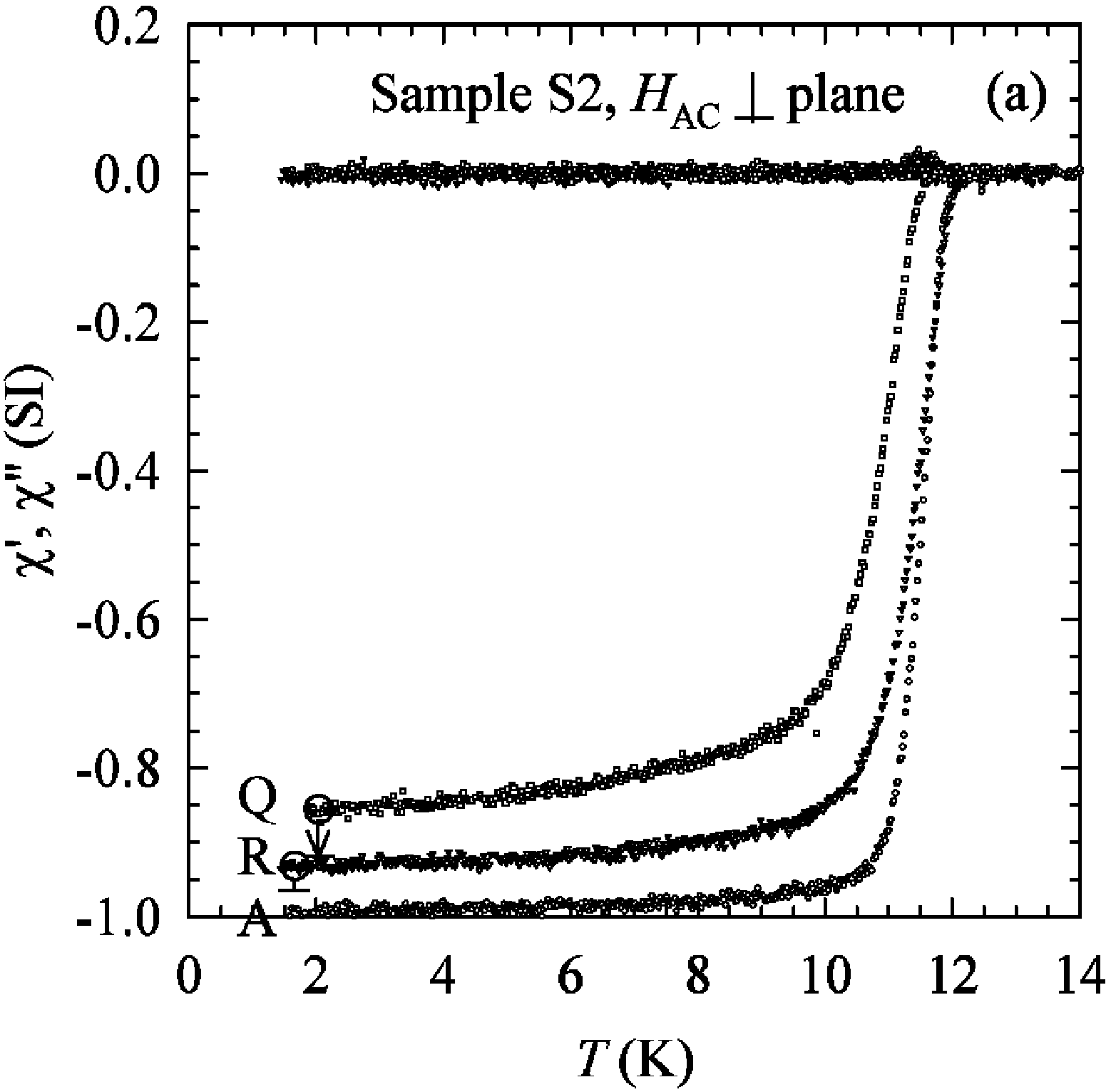}
\centering\includegraphics[clip,scale=0.50]{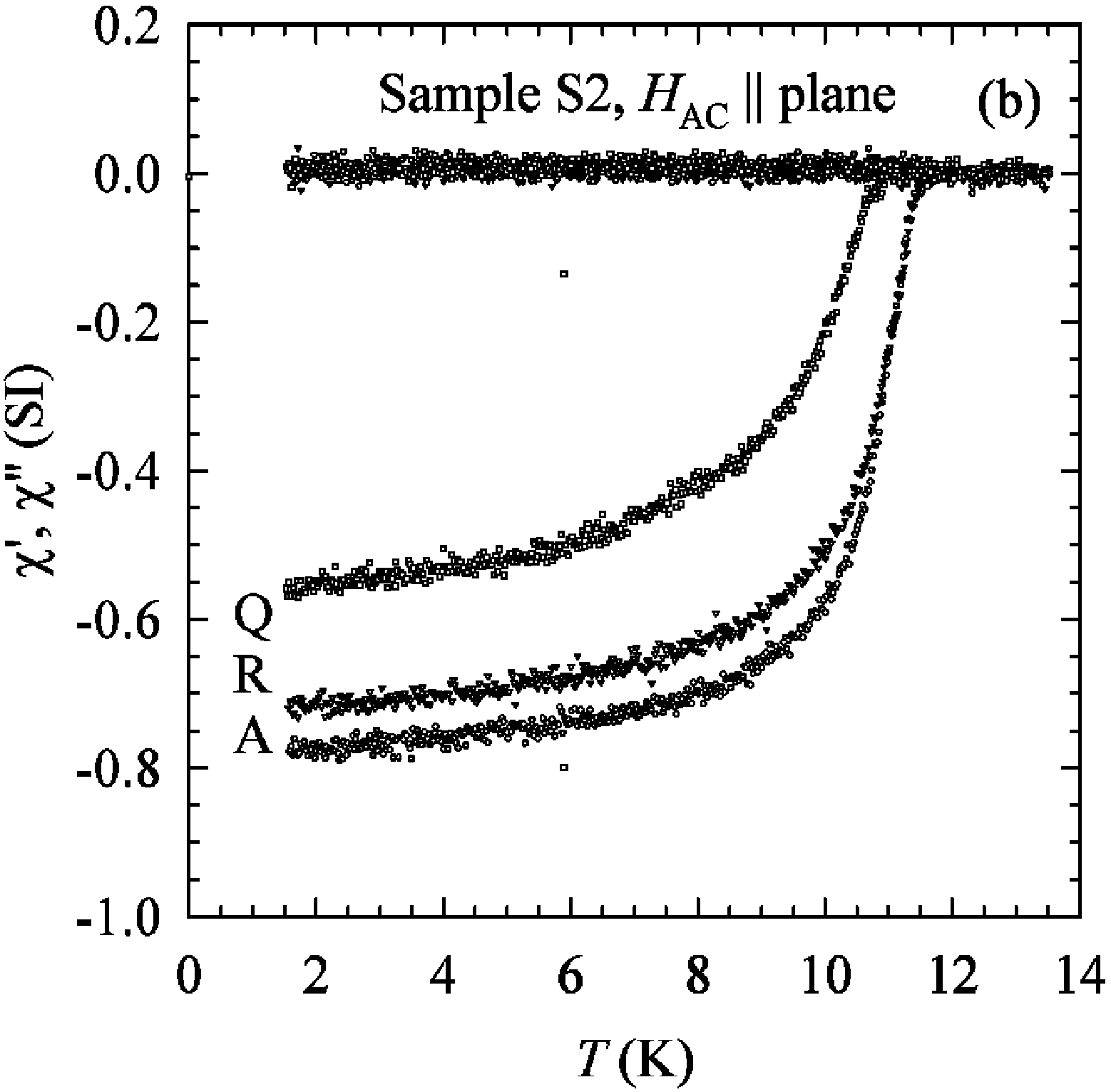}
\caption{Sample~S2: Real and imaginary part of susceptibility for Annealed (A),
Relaxed (R) and Quenched (Q) state in (a)~$H_{\text{AC}} \perp$ 
plane and (b)~$H_{\text{AC}} \parallel$ plane geometry.  The arrows 
in (a) illustrate the upper limit of the systematic error.}
\label{cSch}
\end{figure}
Susceptibility data obtained for sample~S2 for three different cooling 
rates are presented in Fig.~\ref{cSch}.  The first feature, reflecting a
different sample quality, is that the ground state, characterized as 
before by almost complete diamagnetic response, is established in 
the A~state, and not, as in the case of sample~S1, in the R~state.  For a 
purpose of clarity, we will refer to the R~state of sample~S2 as the 
\textit{intermediate state}\footnote{This state should not, by any 
means, be identified with the historical notion of the intermediate
state in superconductors.\cite{inter}}.  The anisotropy of the susceptibility in 
the ground state is somewhat larger for sample~S2 than for sample~S1.  
The second feature is in that the cooling rate effect on the susceptibility 
value is much smaller for sample~S2 compared with the effects obtained for 
sample~S1 (see Fig.~\ref{cKan}).

Finally, we point out that the $\chi''(T)$ component was negligible,
indicating clearly that the measured samples were in the Meissner state.

\subsection{Penetration depth}
\label{lam}

\begin{figure}
\centering\includegraphics[clip,scale=0.40]{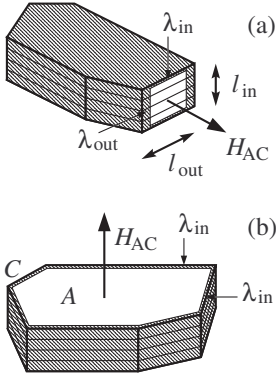}
\caption{Simplified depiction of the sample in the magnetic 
field (a) $H_{\text{AC}} \parallel$ plane and (b) $H_{\text{AC}} \perp$ 
plane.  Shaded parts of the sample represent the
volume penetrated by the magnetic field.  Condition 
$\lambda_{\text{out}} / \lambda_{\text{in}} \gg l_{\text{in}} / l_{\text{out}}$
in (a) ensures that $\lambda_{\text{in}}$ can be neglected in
the analysis\cite{Pinteric:PRB:Br}.}
\label{lama}
\end{figure}
Further, we analyse susceptibility data for hereabove described single 
crystals of different quality in order to get the penetration depth 
temperature behavior and the values at zero temperature.  In 
$H_{\text{AC}} \parallel$ plane geometry we have already argued that the in-plane
penetration depth $\lambda_{\text{in}}$ can be neglected in analysis
and that the out-of-plane penetration depth $\lambda_{\text{out}}$ can be 
obtained from the susceptibility data using the formula for a thin 
superconducting plate in a parallel field\cite{Pinteric:PRB:Br} 
(see Fig.~\ref{lama}(a))
\begin{equation}
1+\chi'=\frac {2\lambda_{\text{out}}} {l_{\text{out}}} \tanh \left(\frac {l_{\text{out}}} {2\lambda_{\text{out}}} \right)
\label{lam1}
\end{equation}

Now we address in more detail $H_{\text{AC}} \perp$ plane geometry.  
The magnetic field is strictly perpendicular to the conducting 
planes, so is the responding magnetization that expels it
out of the bulk.  The resultant circulating supercurrents will 
therefore flow within planes, which will only give contribution to 
$\lambda_{\text{in}}$ (see Fig.~\ref{lama}(b)).  
In order to obtain an appropriate formula for the analysis,
we start with the generalization of Eq.~(\ref{lam1}).  We take
into account that the ratio between the doubled penetration depth 
$2 \lambda_{\text{out}}$ and the sample width in the direction of 
field penetration $l_{\text{out}}$ is the ratio between the 
volume penetrated by the magnetic field $V_{\text{P}}$ and the whole 
sample volume $V$.  Therefore,
\begin{equation}
1 + \chi' = \frac{V_{\text{P}}}{V} \tanh \frac{V}{V_{\text{P}}}.
\label{lam2}
\end{equation}

We note that, if the sample has the shape of the platelet, the 
magnetic field penetrates along the whole edge of the face.  We
can rewrite Eq.~(\ref{lam2}) as
\begin{equation}
1 + \chi' = \frac{C \lambda_{\text{in}} }{A} \tanh \frac{A}{C \lambda_{\text{in}}},
\label{lam3}
\end{equation}
where $C$ is circumference, $A$ area of the platelet face and $\lambda_{\text{in}}$
the in-plane penetration depth (see Fig.~\ref{lama}(b)).

\begin{figure}
\centering\includegraphics[clip,scale=0.50]{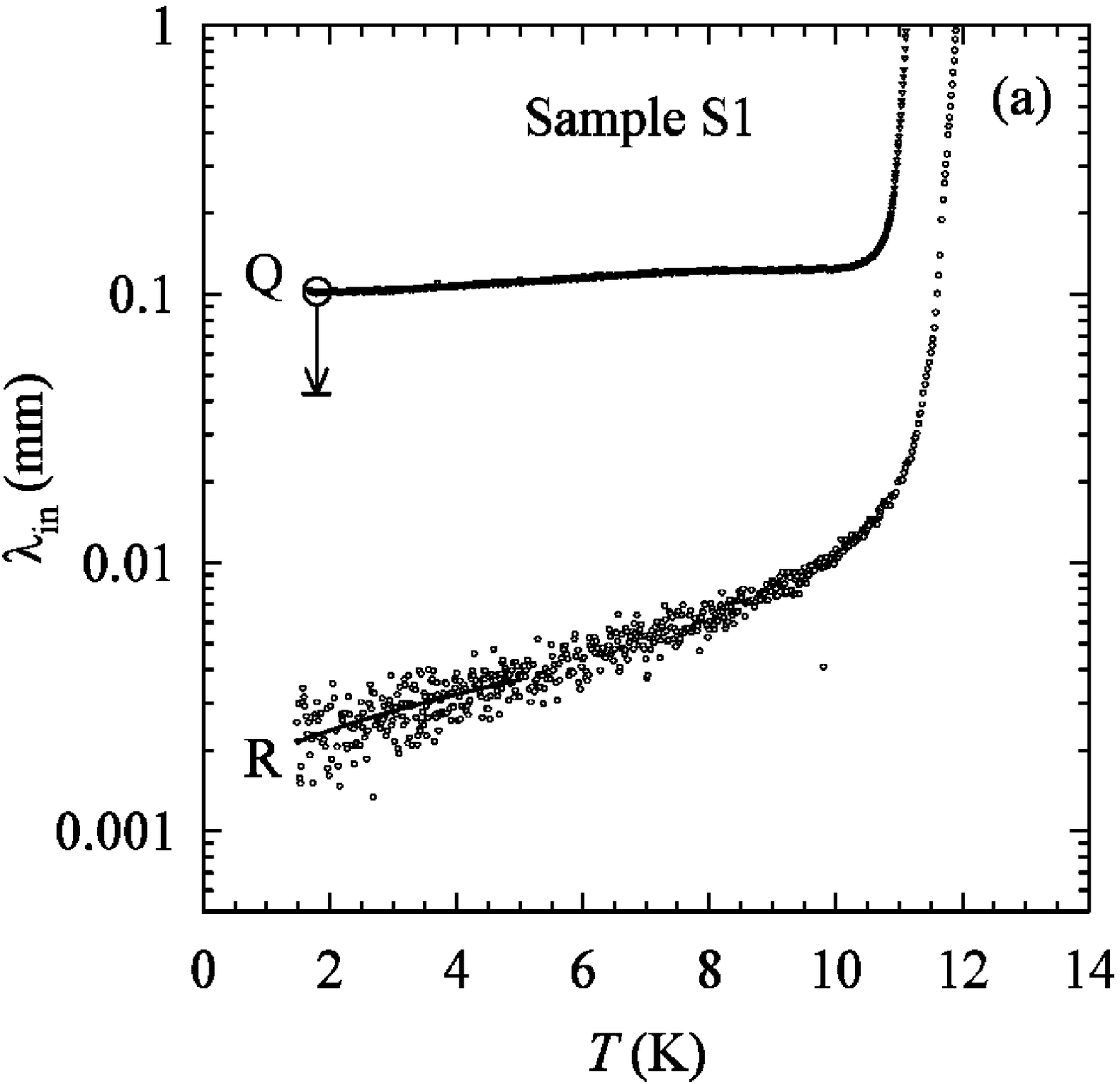}
\centering\includegraphics[clip,scale=0.50]{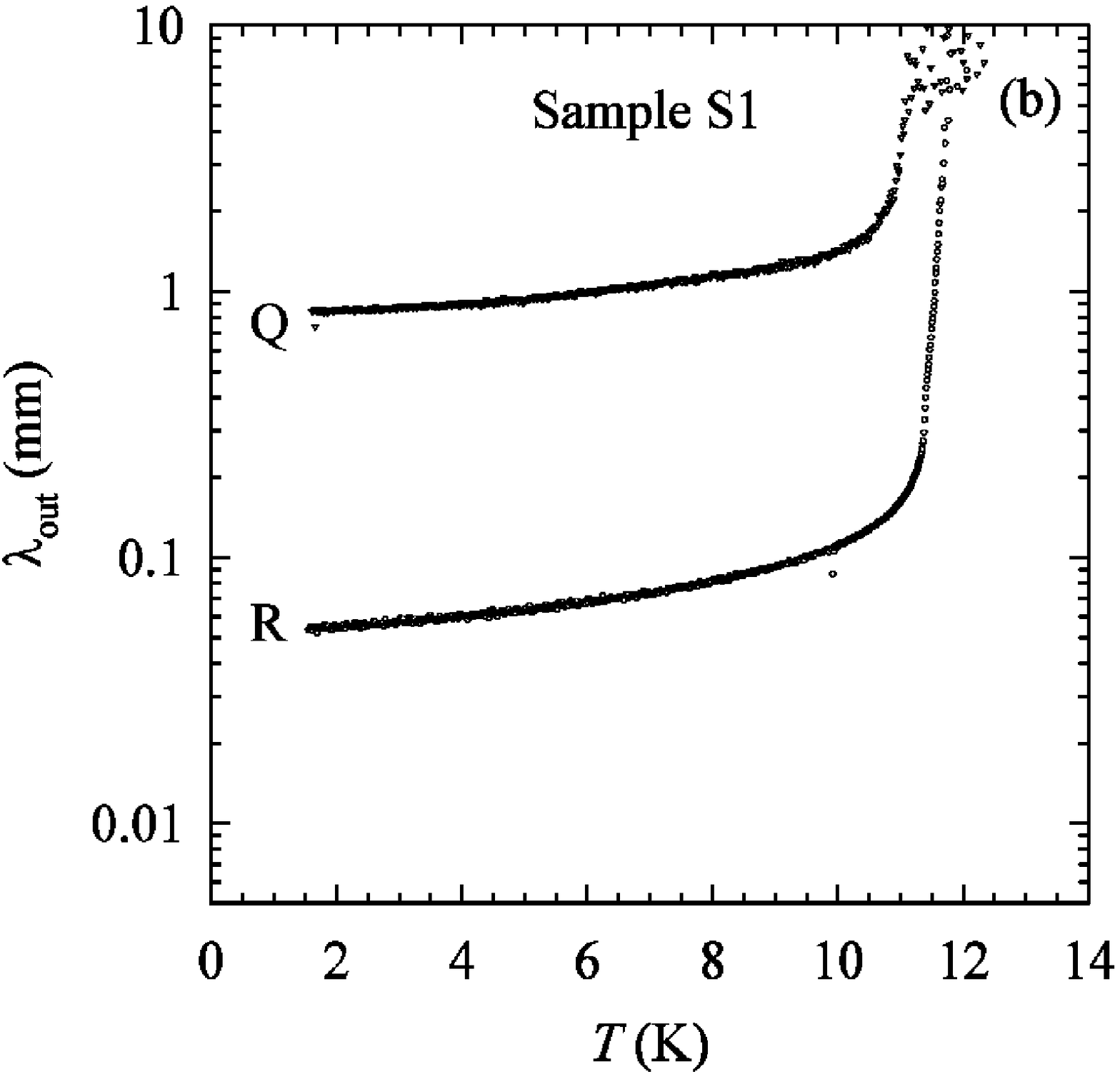}
\caption{Sample~S1: (a) In-plane and (b) out-of-plane 
penetration depth for Relaxed (R) and Quenched (Q) state.  Full 
lines represent the fit to the power law behavior, while the 
arrow in (a) illustrate the upper limit of the systematic error.}
\label{lKan}
\end{figure}
\begin{figure}
\centering\includegraphics[clip,scale=0.50]{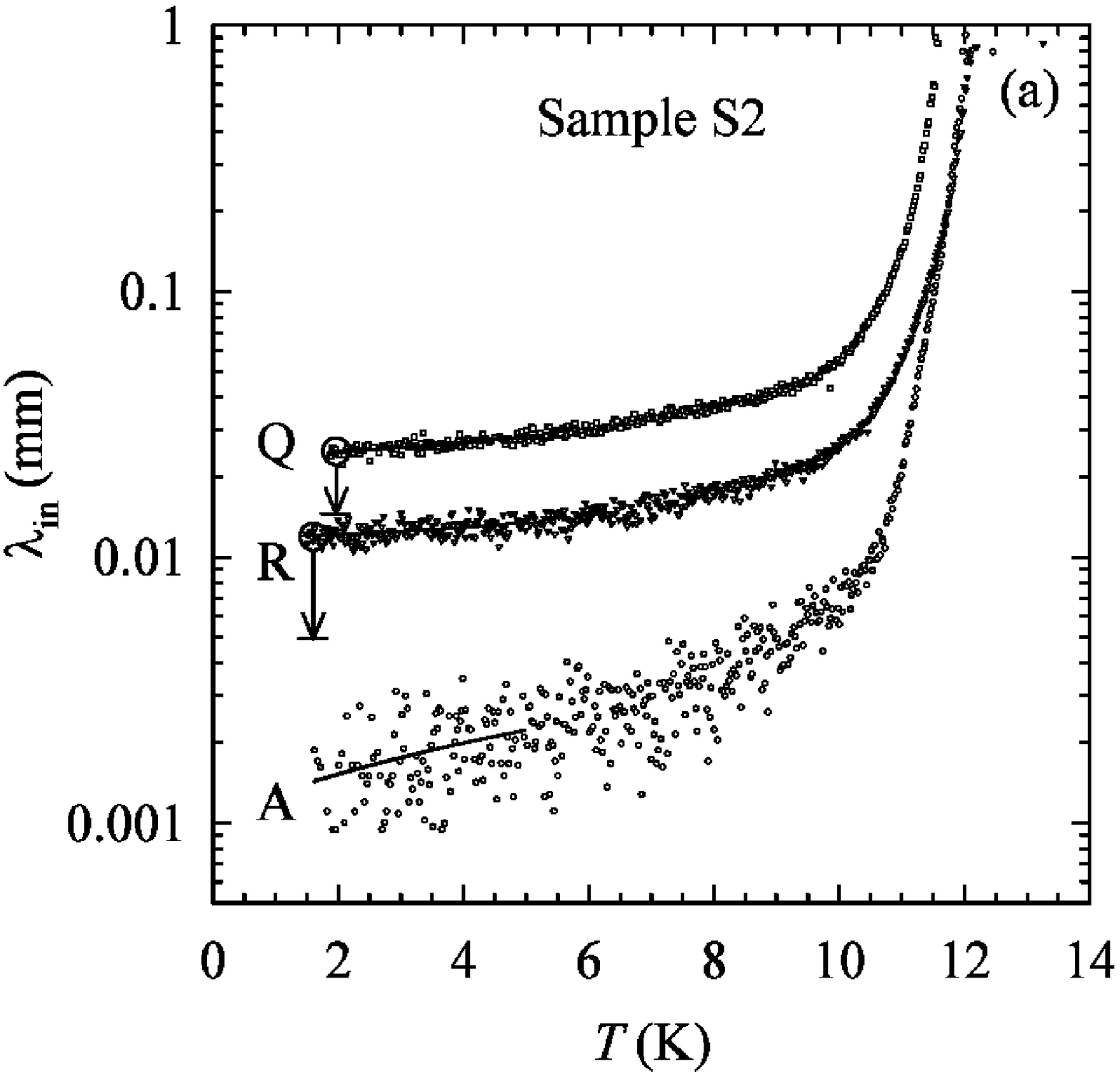}
\centering\includegraphics[clip,scale=0.50]{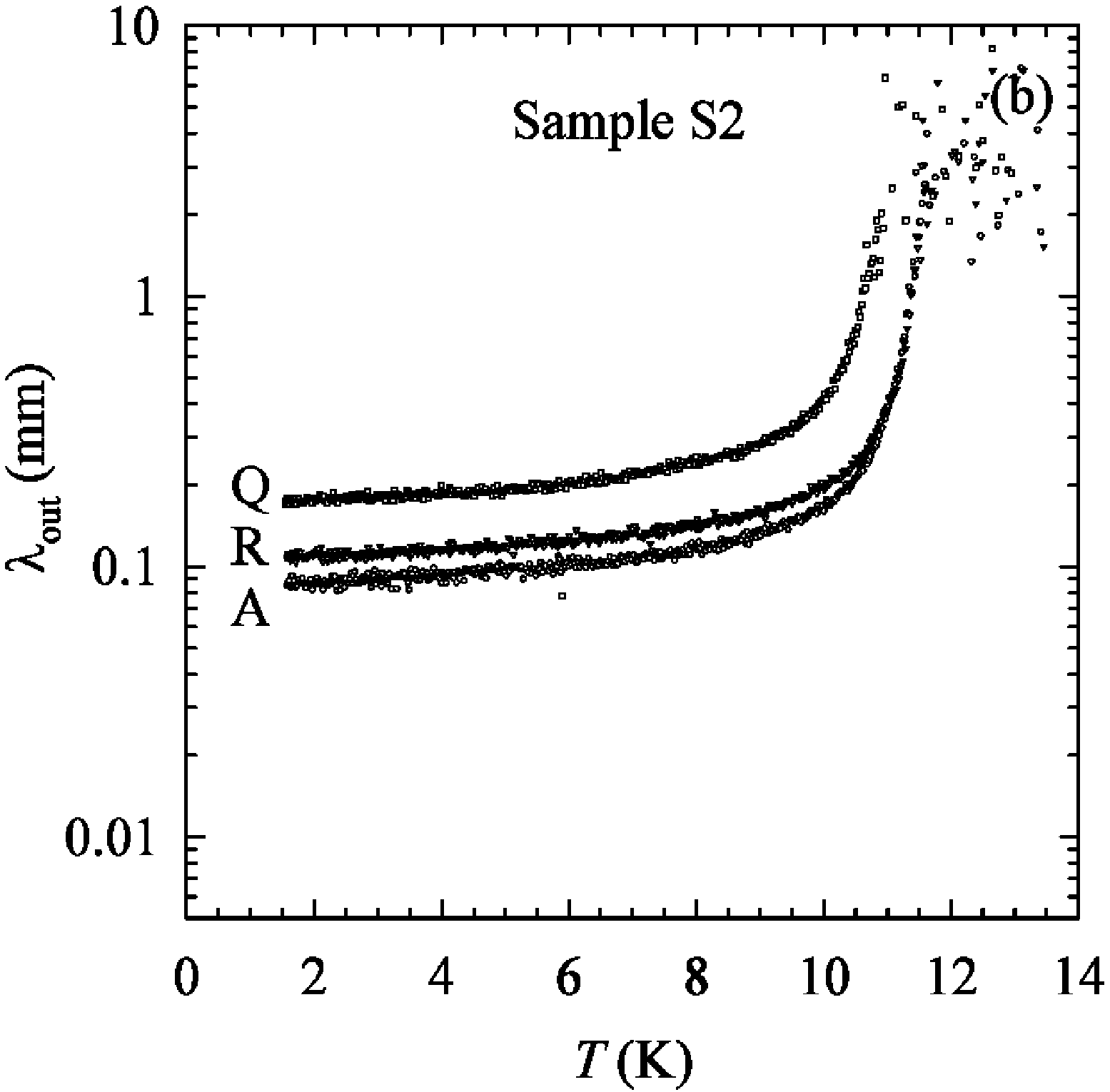}
\caption{Sample~S2: (a) In-plane and (b) out-of-plane 
penetration depth for Annealed (A), Relaxed (R) and Quenched (Q)
state.  Full lines represent the fit to the power law behavior, 
while the arrows in (a) illustrate the upper limit of the 
systematic error.}
\label{lSch}
\end{figure}
Our second important result concerns the temperature dependence
of $\lambda_{\text{in}}$ and $\lambda_{\text{out}}$ in the ground
state as a function of the synthesis procedure.  In Figs.~\ref{lKan} 
and~\ref{lSch} we show the influence of cooling rate on the 
anisotropic penetration depth as a function of the sample quality. 
Note that in the ground state of both samples~S1 and~S2 the temperature
dependence of $\lambda_{\text{in}}$ and $\lambda_{\text{out}}$ 
at temperatures below about 5~K is well described by the $T$ and 
$T^{2}$ law, respectively.  The upper bound of the fit range (5~K) is given
by the general requirement that the genuine low-temperature
behaviour of the penetration depth is strictly obeyed only far from
the critical region close to $T_{\text{C}}$.
The full lines correspond to the calculated fit to the power 
law behavior in the temperature range 1.6~K~$< T <$~5~K
\begin{eqnarray}
\lambda_{\text{in}} = k_{\text{in}} \left( \frac {T} {T_{C}} \right) + \lambda_{\text{in}}(0)
\label{lam4}\\
\lambda_{\text{out}} = k_{\text{out}} \left( \frac {T} {T_{C}} \right)^{2} + \lambda_{\text{out}}(0).
\label{lam5}
\end{eqnarray}
We get $k_{\text{in}} = 5.2~\mu$m, $\lambda_{\text{in}}(0) = 1.5 \pm 0.5 \mu$m,
$k_{\text{out}} = 56~\mu$m, $\lambda_{\text{out}}(0) = 53 \pm 10 \mu$m and
$k_{\text{in}} = 2.8~\mu$m, $\lambda_{\text{in}}(0) = 1.1 \pm 0.4 \mu$m,
$k_{\text{out}} = 69~\mu$m, $\lambda_{\text{out}}(0) = 85 \pm 10 \mu$m
for~S1 and~S2, respectively.  The penetration depth values 
at 0~K, $\lambda(0)$, observed in the ground states of both 
samples, are in a very good accordance with values for the 
penetration depths given in the literature \cite{Pinteric:PRB:Br,Carrington:PRL:ET,%
Lang:SCR:org,Mansky:PRB:NCS,Tea:PHC:Br,Dressel:PRB:ET,Lee:PRL:NCS,Achkir:PRB:NCS}.

A special attention should be given to the relative change of 
the penetration depth at low temperatures,
$\eta(T) = (\lambda(T)-\lambda(0))/\lambda(0)$.  We denote 
the deviation of $\lambda_{\text{in}}(T)$ and $\lambda_{\text{out}}(T)$
from their values at 0~K, in units of $\lambda(0)$, as 
$\eta_{\text{in}}(T)$ and $\eta_{\text{out}}(T)$, respectively.
We find that $\eta_{\text{in}}(5 \textrm{K}) = 1.4$, 
$\eta_{\text{out}}(5 \textrm{K}) = 0.19$ and 
$\eta_{\text{in}}(5 \textrm{K}) = 1.1$, $\eta_{\text{out}}(5 \textrm{K}) = 0.15$ 
in the ground state of samples~S1 and~S2, respectively.  
First we note that $\eta$ values are in perfect accordance 
for both samples confirming that they are in the same 
-- ground -- state.  Second, a difference between 
$\eta_{\text{in}}$ and $\eta_{\text{out}}$ values for almost 
an order of magnitude proves a strong anisotropy in physical 
properties between the two orientations.  This also confirms 
our choice for the exponent in the power law in 
Eqs.~(\ref{lam4}) and~(\ref{lam5}).

Finally, our third important result concerns the intermediate
state (R~state in sample~S2), for which we get
$\eta_{\text{in}}(5 \textrm{K}) = 0.13$, $\eta_{\text{out}}(5 \textrm{K}) = 0.10$.
Unlike for the ground state, the temperature dependences of 
$\lambda_{\text{in}}$ and $\lambda_{\text{out}}$ are so similar 
that they cannot be any longer described by different power 
laws as was the case in the ground state.  The obvious solution 
is to try to fit both penetration depths to the $T^2$ 
power law behavior
\begin{eqnarray}
\lambda_{\text{in}} = k_{\text{in}} \left( \frac {T} {T_{C}} \right)^{2} + \lambda_{\text{in}}(0)
\label{lam6}\\
\lambda_{\text{out}} = k_{\text{out}} \left( \frac {T} {T_{C}} \right)^{2} + \lambda_{\text{out}}(0),
\label{lam7}
\end{eqnarray}
which suggests a $d$-wave superconductor with impurities.  We get 
$k_{\text{in}} = 9.2~\mu$m, $\lambda_{\text{in}}(0) = 12 \pm 6 \mu$m, 
and $k_{\text{out}} = 58~\mu$m, $\lambda_{\text{out}}(0) = 110 \pm 20 \mu$m.  
We point out that the fit to the $s$-wave model describe our data 
almost equally well (see Section \ref{R-sw}).  On the other hand, the $s$-wave 
model fails completely for the penetration depth temperature dependences in the ground state 
of both samples~S1 and~S2.

Now we comment penetration depth results for Q~state.  We find that 
$\eta_{\text{in}}(5 \textrm{K}) = 0.11$, $\eta_{\text{out}}(5 \textrm{K}) = 0.12$ 
and $\eta_{\text{in}}(5 \textrm{K}) = 0.16$, $\eta_{\text{out}}(5 \textrm{K}) = 0.10$ 
for samples~S1 and~S2, respectively.  Here we apply the same arguments 
as in the case of the intermediate state and fit both $\lambda_{\text{in}}$ 
and $\lambda_{\text{out}}$ to the $T^2$ power law behavior (Eqs.~(\ref{lam6}) 
and~(\ref{lam7})).  We get
$\lambda_{\text{in}}(0) = 100 \pm 50 \mu$m, $\lambda_{\text{out}}(0) = 830 \pm 100 \mu$m,
and $\lambda_{\text{in}}(0) = 24 \pm 12 \mu$m, $\lambda_{\text{out}}(0) = 170 \pm 20 \mu$m 
for sample~S1 and~S2, respectively.  The fact that 
these fits describe the penetration depth data well again suggests 
a $d$-wave superconductor with impurities.
Further, it should be noted that (i)~$\lambda(0)$ values are larger for
Q~state than for the intermediate state, suggesting a larger disorder in 
the former state, (ii)~$\lambda(0)$ values for sample~S1 are significantly 
larger than ones for~S2, suggesting significantly larger disorder in 
the former sample in Q~state.  Finally, the result that $\lambda_{\text{out}}(0)$
for sample~S1 is close to the crystal size indicates that the bulk 
superconductivity is not established, allowing us to define the boundary
between bulk and nonbulk SC at $\chi' = -0.7$ for $H_{\text{AC}} \perp$~plane.

\begin{table}[htb]
\centering
\begin{tabular}{|c|c|c|c|c|c|c|}
\hline
State & $\lambda_{\text{in}}(0)$ & $\lambda_{\text{out}}(0)$ & $k_{\text{in}}$ & $k_{\text{out}}$ & $\eta_{\text{in}}(5 \textrm{K})$ & $\eta_{\text{out}}(5 \textrm{K})$ \\
 & ($\mu$m) & ($\mu$m) & ($\mu$m) & ($\mu$m) & & \\
\hline
\multicolumn{7}{|c|}{Sample S1} \\
\hline
R & $1.5 \pm 0.5$ & $53 \pm 10$ & 5.2 & 56 & 1.4 & 0.19 \\
Q & $100 \pm 50$ & $830 \pm 100$ & 60 & 480 & 0.11 & 0.12 \\
\hline
\multicolumn{7}{|c|}{Sample S2} \\
\hline
A & $1.1 \pm 0.4$ & $85 \pm 10$ & 2.8 & 69 & 1.1 & 0.15 \\
R & $12 \pm 6$ & $110 \pm 20$ & 9.2 & 58 & 0.13 & 0.10 \\
Q & $24 \pm 12$ & $170 \pm 20$ & 22 & 88 & 0.16 & 0.10 \\
\hline
\end{tabular}
\caption{Penetration depth properties, as defined in Text, for
(i) Sample~S1 in Relaxed (ground) and Quenched state, (ii) Sample~S2
in Annealed (ground), Relaxed (intermediate) and Quenched state.}
\label{resume}
\end{table}

\subsection{Superfluid density}
\label{rho}

In the following, we address the temperature dependance 
of the superfluid density in order to get information on the 
symmetry of the superconducting state.  We construct the 
in-plane superfluid density $\rho_{\text{s,in}}$ and 
out-of-plane superfluid density $\rho_{\text{s,out}}$ as 
\begin{eqnarray}
\rho_{\text{s,in}} =  \left(\frac{\lambda_{\text{in}}(0)} {\lambda_{\text{in}}(T)} \right)^{2} \\
\rho_{\text{s,out}} = \left(\frac{\lambda_{\text{out}}(0)}{\lambda_{\text{out}}(T)}\right)^{2}
\end{eqnarray}

\begin{figure}
\centering\includegraphics[clip,scale=0.50]{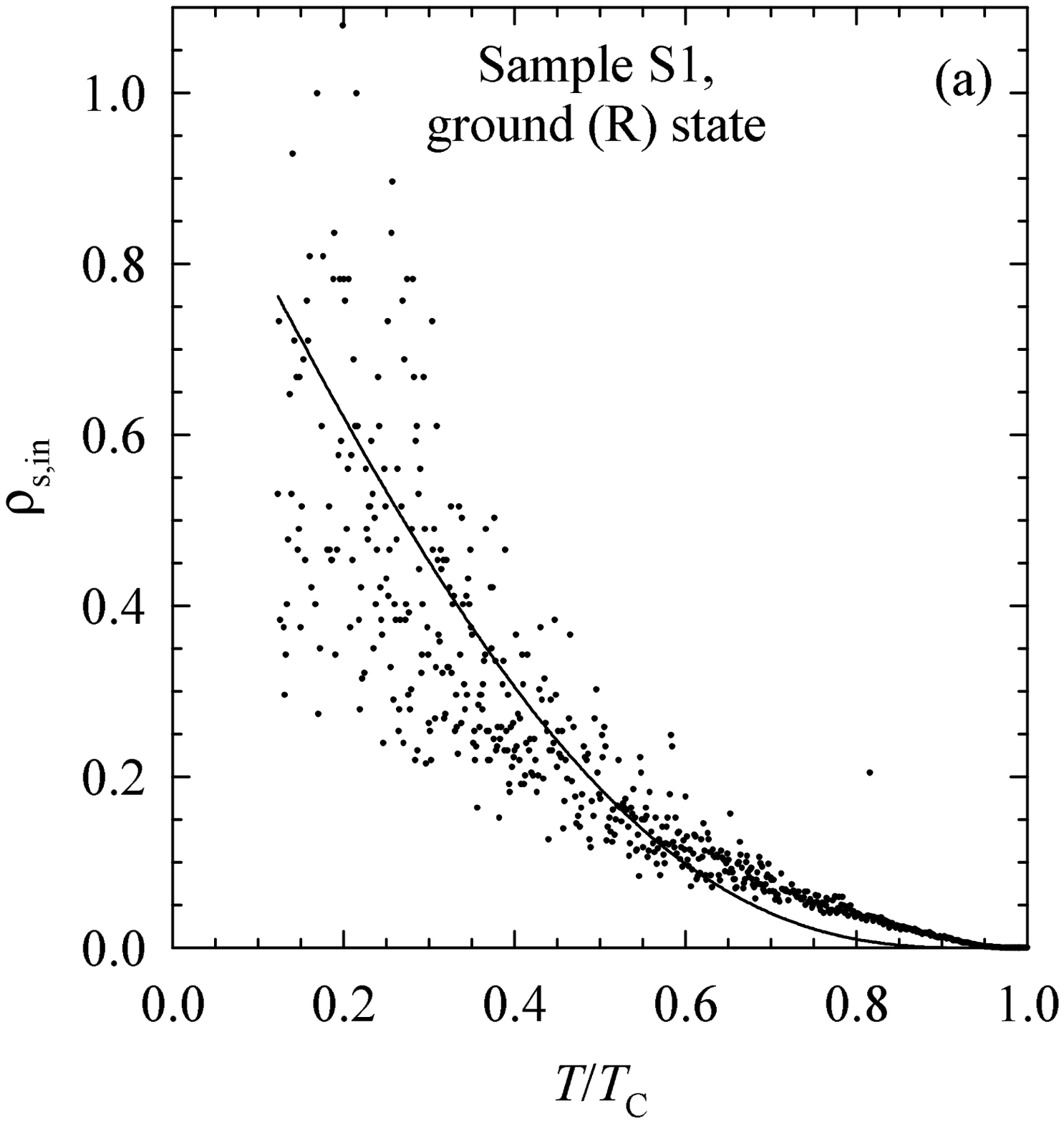}
\centering\includegraphics[clip,scale=0.50]{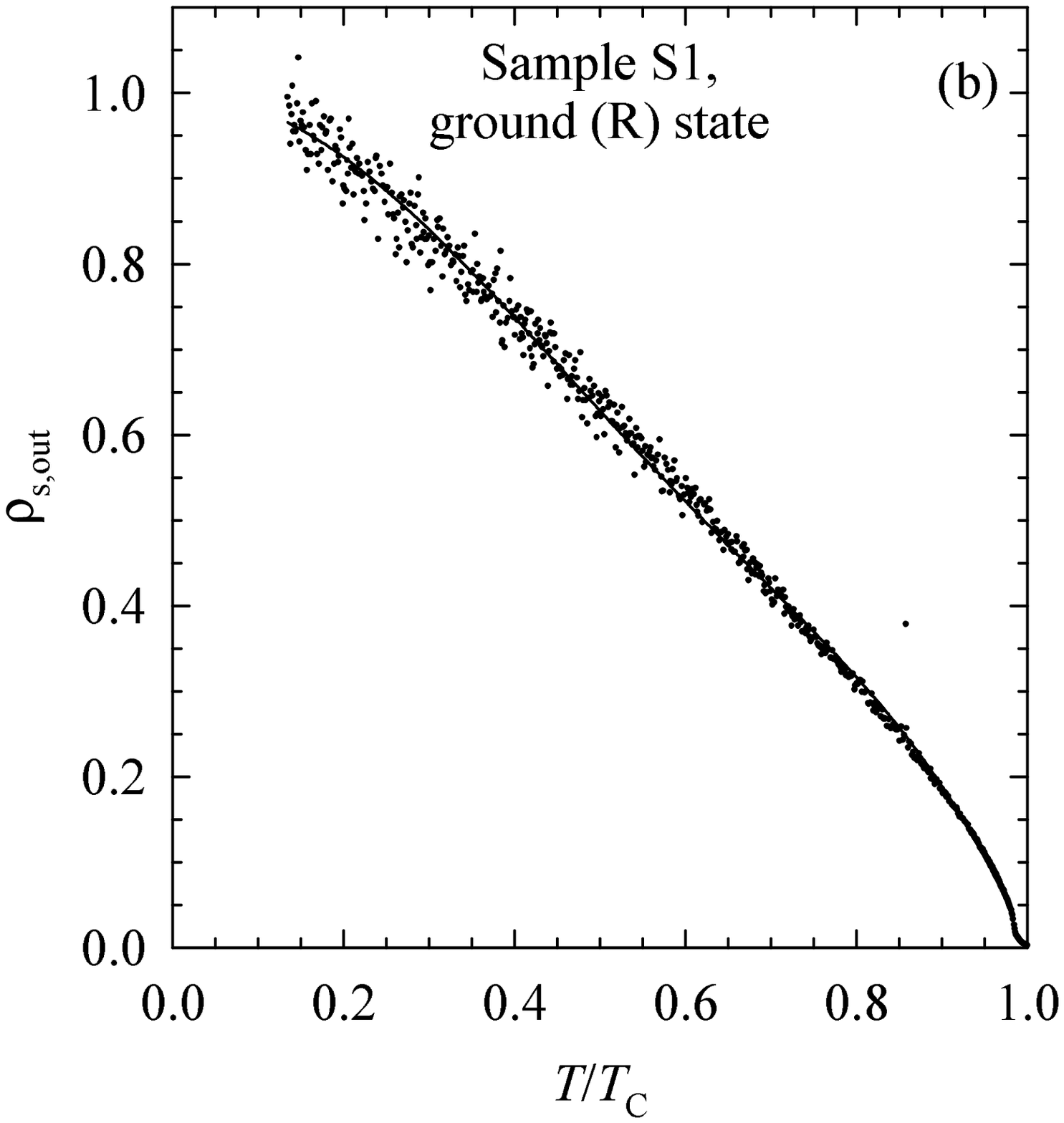}
\caption{Sample~S1: (a) In-plane and (b) out-of-plane 
superfluid density for the ground (R) state.  Solid line 
is a fit to the polinomial expression.  A large noise in
$\rho_{\text{s,in}}$ is due to small values of 
$\lambda_{\text{in}}(0)$ (see Text).}
\label{rKanR}
\end{figure}
\begin{figure}
\centering\includegraphics[clip,scale=0.50]{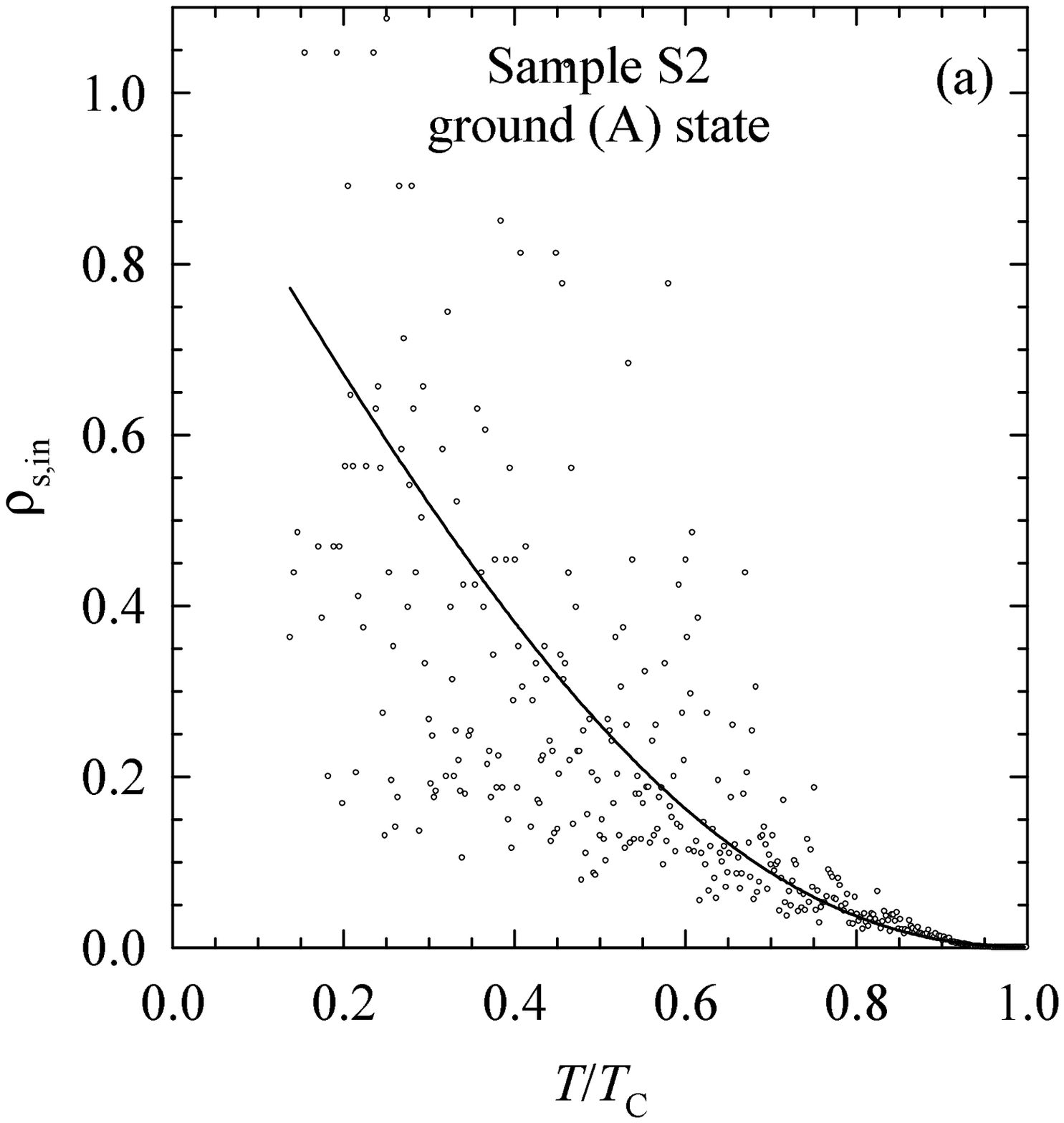}
\centering\includegraphics[clip,scale=0.50]{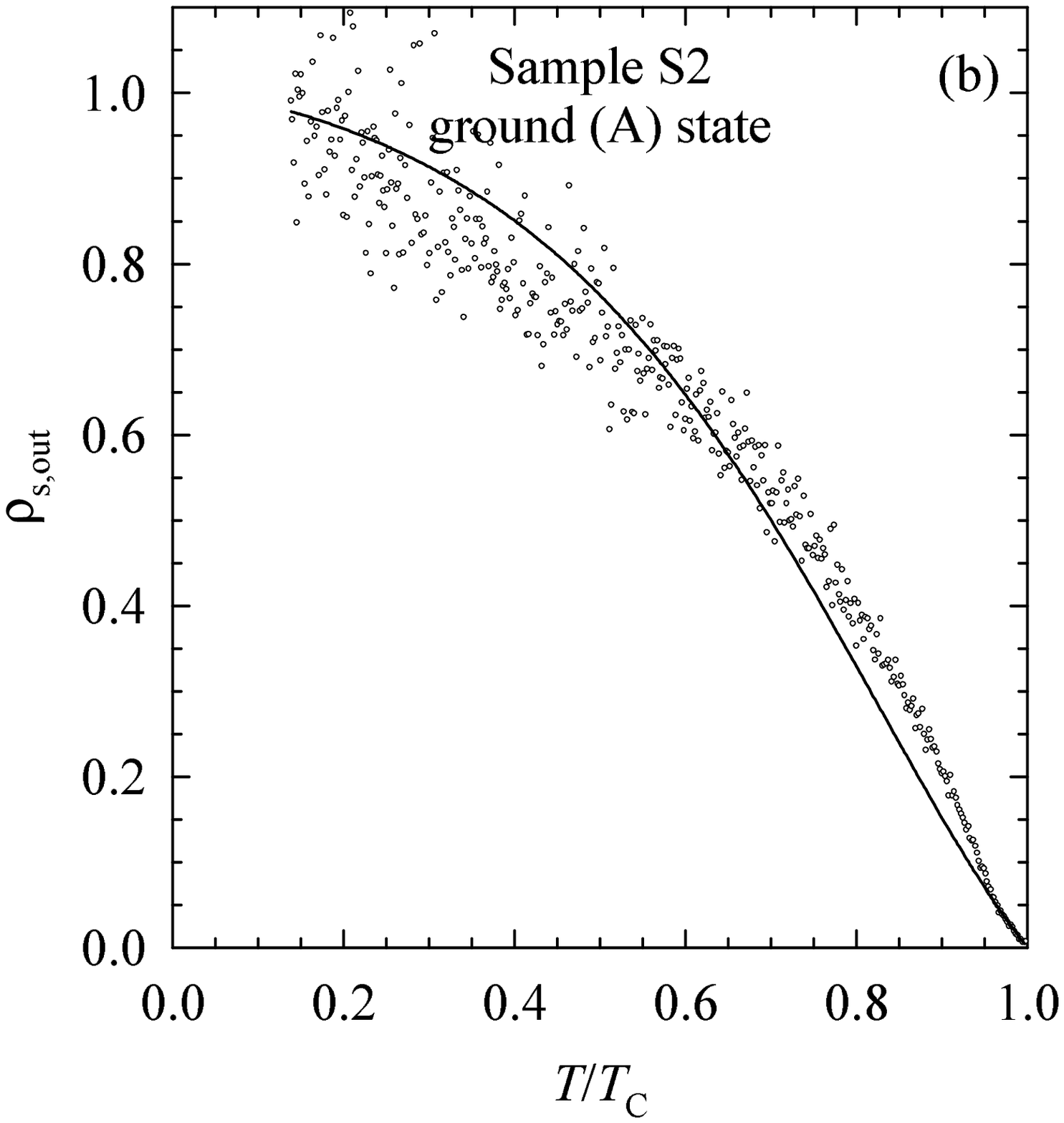}
\caption{Sample~S2: (a) In-plane and (b) out-of-plane 
superfluid density for the ground (A) state.  Solid line 
is a fit to the polinomial expression.  A large noise in
$\rho_{\text{s,in}}$ is due to small values of 
$\lambda_{\text{in}}(0)$ (see Text).}
\label{rSchA}
\end{figure}
$\rho_{\text{s,in}}$ and $\rho_{\text{s,out}}$ 
for the ground state of sample~S1 (established 
in the R~state) and for the ground state of sample~S2 (established in 
the A~state) as a function of reduced temperature $t = T/T_{\text{C}}$ 
are displayed in Figs.~\ref{rKanR} and~\ref{rSchA}, respectively.  
There is a strong resemblance in the behavior found for both samples.  
Note that the leading terms, which describe the low temperature 
behavior, are $T$~and $T^2$~term for $\rho_{\text{s,in}}$ and 
$\rho_{\text{s,out}}$, respectively\cite{Pinteric:PRB:Br}.  This is 
to be expected, because $T$ and $T^2$ terms describe the low temperature 
behavior of the $\lambda_{\text{in}}$ and $\lambda_{\text{out}}$ in 
the ground state.  If we expand the leading term to the full polynomial, 
in order to fit the superfluid density data in the whole temperature 
region below $T_{\text{C}}$, we finally obtain for sample~S1
\begin{eqnarray}
\rho_{\text{s,in}} = 1 - 1.95 t + 1.45 t^3 - 0.09 t^4 - 0.41 t^5\label{rsin1}\\
\rho_{\text{s,out}} = 1 - 1.88 t^2 - 0.73 t^3 + 4.47 t^4 - 2.86 t^5,
\end{eqnarray}
and for sample~S2
\begin{eqnarray}
\rho_{\text{s,in}} = 1 - 1.68 t + 0.78 t^3 + 0.16 t^4 - 0.26 t^5\label{rsin2}\\
\rho_{\text{s,out}} = 1 - 1.45 t^2 + 2.98 t^3 - 5.38 t^4 + 2.84 t^5.
\end{eqnarray}
Taking into account a relatively large experimental error in the 
penetration depth values (see Figs.~\ref{lKan} and~\ref{lSch}), the 
leading coefficient values might be considered to be almost the same.  
In addition to the systematic error, there is also a noise, which
is a mere consequence of the fact that $\rho_{\text{s,in}}$ is 
calculated according to the expression 
$\rho_{\text{s,in}}(T) = (\lambda_{\text{in}}(0)/\lambda_{\text{in}}(T))^{2}$,
so that the absolute noise in $\rho_{\text{s,in}}$ is proportional to 
the relative noise in the penetration depth data.  That implies a larger 
noise for smaller values of $\lambda_{\text{in}}(0)$, which becomes 
substantial for $\lambda_{\text{in}}(0)$ of the order of $1 \mu$m.
Finally, we point out that the 
shape of curves for the in-plane and the out-of-plane 
penetration depth in the ground state are qualitatively 
different from the $s$-wave dependence.

\begin{figure}
\centering\includegraphics[clip,scale=0.50]{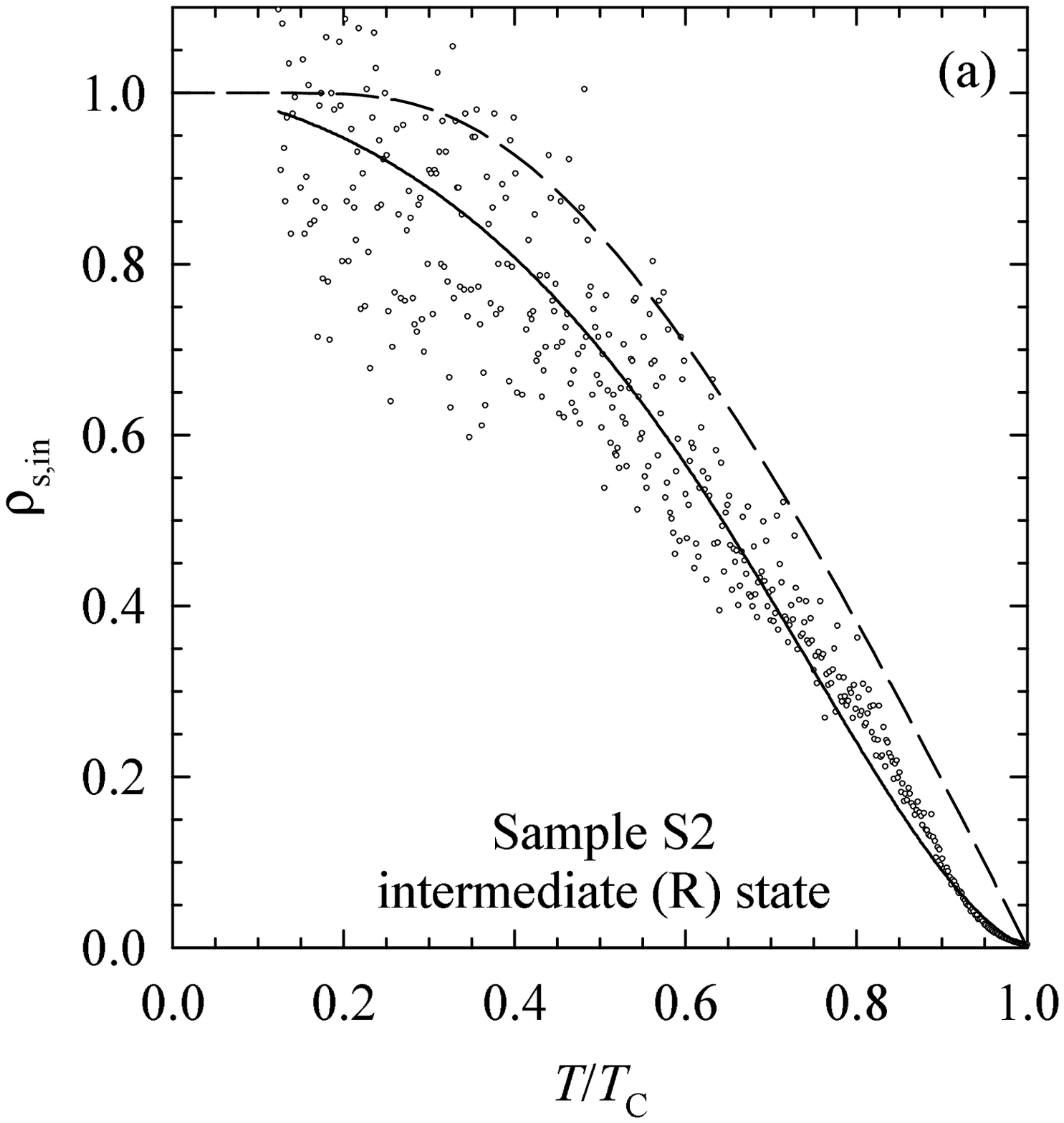}
\centering\includegraphics[clip,scale=0.50]{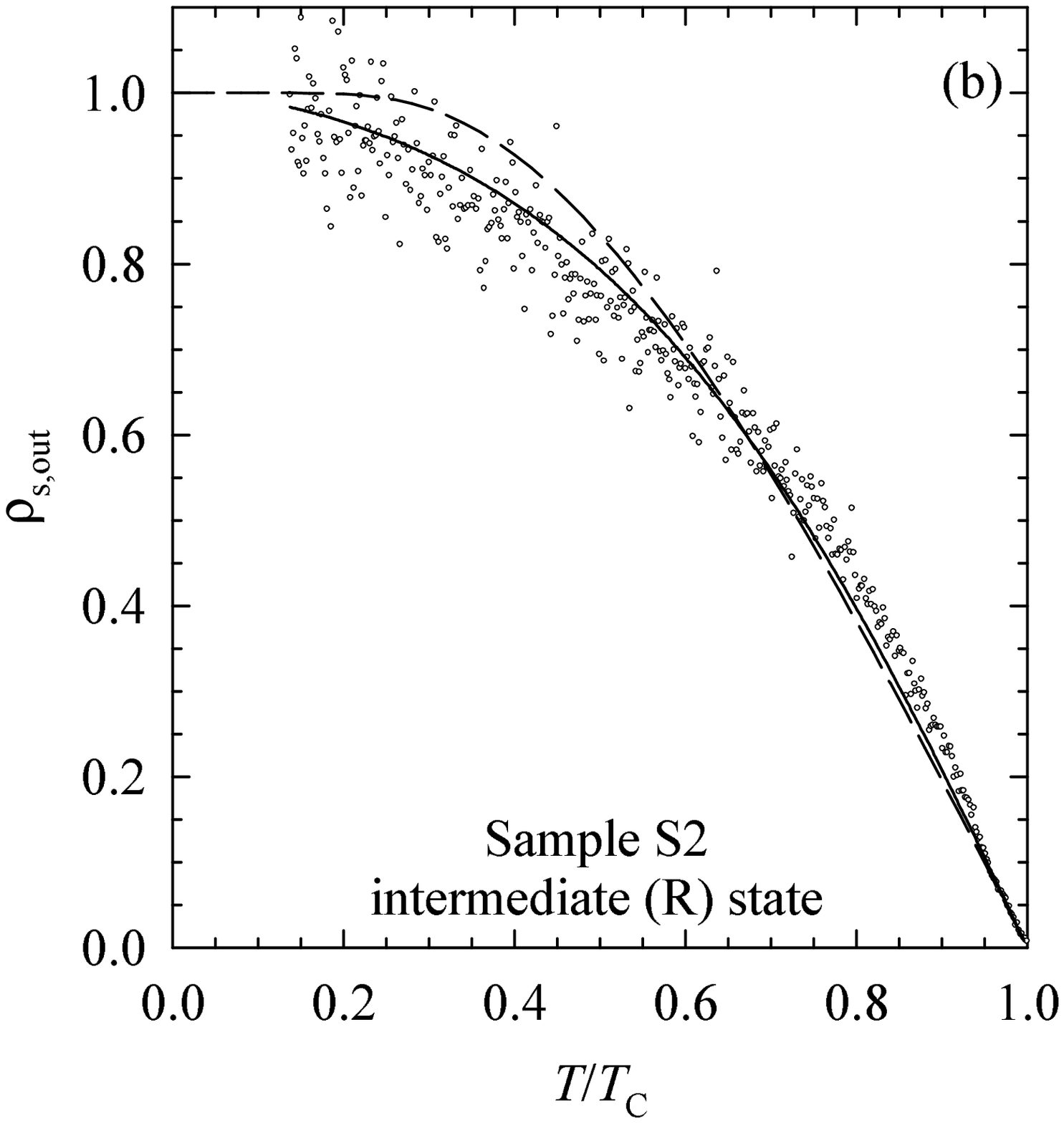}
\caption{Sample S2: (a) In-plane and (b) out-of-plane 
superfluid density for the intermediate (R) state.  Solid line is a fit
to the polinomial expression, and dashed line presents the $s$-wave 
model (see Text).}
\label{rSchR}
\end{figure}
The behavior for the in-plane $\rho_{\text{s,in}}$ and 
out-of-plane superfluid density $\rho_{\text{s,out}}$ for 
the intermediate state of sample~S2 (established in
R~state) is displayed in Fig.~\ref{rSchR}.  
Here, $\rho_{\text{s,in}}$ data are insensitive to the
systematic error at low temperatures, since the correction for
$\lambda_{\text{in}}$ drops out from the expression for 
$\rho_{\text{s,in}}$.  Note that the leading term,
describing the low temperature behavior, is $T^2$ term
for both the in-plane and out-of-plane superfluid density:
\begin{eqnarray}
\rho_{\text{s,in}} = 1 - 1.69 t^2 + 2.77 t^3 - 5.05 t^4 + 2.97 t^5\label{rsin3}\\
\rho_{\text{s,out}} = 1 - 1.02 t^2 + 1.34 t^3 - 2.50 t^4 + 1.18 t^5
\end{eqnarray}
In order to demonstrate the fact that the superfluid density 
behavior in the intermediate state is also rather close to the 
dependence expected for the $s$-wave order parameter behavior, 
the $s$-wave model dependence is added as a dashed line for both 
orientations.\label{R-sw}

\section{Discussion}

We start discussion by pointing out that the well-defined, ground 
state properties - complex susceptibility, penetration depth and 
superfluid density - were essentially reproducible for all studied 
single crystals from both syntheses S1 and S2.  Cooling rate dependent 
effects were also reproducible, but the observed behavior was the same 
only for single crystals from the same synthesis, while it differed
significantly from the observed behavior for single crystals from the 
other synthesis.

First, we would like to comment on the anisotropy in $T_{\text{C}}$.  As 
in the previously published paper\cite{Pinteric:PRB:Br}, we have 
established that $\Delta T_{\text{C}} = 0.5$~K for $H_{\text{AC}} \perp$ 
and $H_{\text{AC}} \parallel$ plane geometry cannot be ascribed to the 
experimental error.  We suggest that this anisotropy might be the
consequence of the fact that the diamagnetic shielding is no longer 
effective for $H_{\text{AC}} \parallel$ plane geometry in the range 
of 0.5~K below $T_{\text{C}}$, which is due to small sample dimensions 
and huge out-of-plane (Josephson-like) penetration depth near 
$T_{\text{C}}$.  However, we point out that 
$T_{\text{C}}$ anisotropy has negligible effects, if any, to our data 
analysis, which is primarily done in the low-temperature region, and 
therefore does not influence the resulting conclusions.

Second, we address the observed differences in cooling rate effects 
between samples~S1 and~S2.  As pointed out earlier, both samples show the 
same behavior in the ground state: almost full diamagnetism for 
$H_{\text{AC}} \perp$ plane, the same temperature dependence and 
zero-temperature value of the in-plane and the out-of-plane penetration 
depth, as well as the same superfluid density temperature dependence.  
However, we note that sample~S2 required a completely different cooling 
procedure with significantly longer time spent in the temperature 
region around 80~K, compared to sample~S1, in order to accomplish the 
ground state.  Moreover, sample~S2 is much less sensitive to the 
cooling rate.  That is, the difference regarding low-temperature 
susceptibility and zero-temperature penetration depth values between 
A~and Q~states for sample~S2 is much smaller than the difference 
between R~and Q~states for sample~S1.  Both features indicate
that (i)~the low-temperature state is critically determined by the time
scale of experiment in the region of glass transition and (ii)~relaxation 
times of ethylene groups in the single crystals originating
from synthesis~S2 are much longer than the ones in the single crystals
originating from synthesis~S1.  When the applied time scale is much
longer than the relaxation time of the ethylene moieties, the
low-temperature state is the ground state.  In contrast, if the 
relaxation time exceeds the time scale of experiment, remnant disorder 
at low temperatures will be substantial and Q~state will be established.
Different relaxation times of ethylene groups might also explain why
the resistivity ratio $RR(75\textrm{K}/T_{\text{min}}$) is much larger for
samples from synthesis~S1 when standard slow cooling rate is applied.  At this
stage, we can only speculate about the possible origin of different 
relaxation times.  The experimental observations that the crystals from 
synthesis~S2 show weak metallic behavior, instead of a semiconducting 
behavior between RT and 100~K observed for samples of synthesis~S1, might
be of the same origin.   Since the RT resistivity values do not 
differ substantially, we propose that the subtle local variations of 
impurity level in nominally pure samples from different syntheses 
might be responsible for the observed differences.

Next we comment on the behavior of the in-plane superfluid
density in the ground state.  The temperature dependence of the 
in-plane superfluid density for the $d$-wave superconducting order 
parameter $\Delta(\vec{k}) = \Delta \cos (2 \phi)$, 
where $\phi$ is the angle between the quasiparticle momentum $\vec{k}$ 
and the $\vec{a}$ axis, within the weak coupling theory is given by 
\cite{Won:PRB:dw,Pinteric:PRB:Br}
\begin{eqnarray}
\rho_{\text{s,in}}(t) &\approx& 1 - 0.6478 t - 0.276 t^{3}
\label{rsint}
\end{eqnarray}
The coefficient $a$ of the leading term $t$ in 
$\rho_{\text{s,in}} = 1 - a t + \ldots$
depends strongly on the ratio of the superconducting transition 
temperature and the zero temperature superconducting order 
parameter.  A comparison of values for $a$ in Eq.~(\ref{rsint}) 
to those in Eqs.~(\ref{rsin1}) and~(\ref{rsin2}) 
suggests that the superconducting order parameter at $T = 0$~K
is much smaller than that predicted by the weak-coupling limit.
As a result, this would also imply that the nodal region, 
which volume is inversely proportional to the angular slope of the 
gap near the node\cite{Xu:PRB:SC} 
$\mu =1/\Delta \cdot d\Delta(\phi)/d\phi|_{\text{node}}$,
occupies a much larger fraction of the phase space at low
temperatures.  We point out that we have reached the same
conclusion already on the basis of previously obtained
results\cite{Pinteric:PRB:Br}.  In the latter case, we did not evaluate 
$\lambda_{\text{in}}(0)$ from our measurements, but rather we 
used the values reported in the literature.  Following 
the same procedure as in Ref.~\onlinecite{Pinteric:PRB:Br}, 
we can again use uncalibrated data from present 
measurements and calculate the deviation of
$\lambda_{\text{in}}$ from the minimum value at the lowest 
attainable temperature 
$\lambda_{\text{in}}(T)-\lambda_{\text{in}}(T_{\text{min}})$\cite{Pinteric:PRB:Br}.
Then, if we take the penetration depth value at 0~K given in the 
literature\cite{Lang:SCR:org,Mansky:PRB:NCS,Tea:PHC:Br,Dressel:PRB:ET},
$\lambda_{\text{in}}(0) \approx 1\mu$m, we get the same result,
that is the leading coefficient is much larger than expected in the 
weak-coupling model.  In addition, we point out that the same 
behavior of $\rho_{\text{in}}$ for 
$\lambda_{\text{in}}(0) \le 1.3 \mu$m was reported by Carrington 
\textit{et al.}\cite{Carrington:PRL:ET}.  They pointed out that 
only at $\lambda_{\text{in}}(0) \ge 1.8 \mu$m the slope
becomes similar to the one reported for high-$T_{\text{C}}$ 
cuprate superconductors and expected in the weak coupling
model.  From our present data, we get 
$\lambda_{\text{in}}(0) \approx 3 \mu$m for the crossover 
in-plane penetration depth value.

\begin{figure}
\centering\includegraphics[clip,scale=0.50]{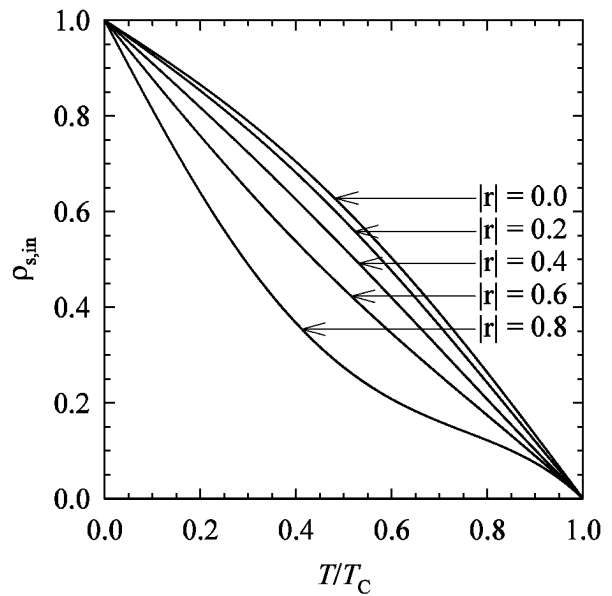}
\caption{Superfluid density in $d+s$-wave model for a few values
of the $s$-wave component parameter~$r$.}
\label{sd}
\end{figure}
One possibility to interpret our results is to consider the mixture 
of $d$-wave and $s$-wave order parameter, which corresponds to the 
superconducting order parameter 
$\Delta(\vec{k}) = \Delta (\cos (2 \phi) + r)$, with $r$ representing 
the $s$-wave component\cite{Won:PHB:SC}.  In this case the leading 
linear coefficient $a$ increases with the increase of $r$ according 
to the expression:
\begin{equation}
a = \frac{2 \ln 2}{2.14 \sqrt{1-r^2}} \exp \left[\frac{2 r^2}{1 + 2r^2}\right]
\end{equation}
The shapes of the superfluid density curves for several values of 
parameter $r$ are given in Fig.~\ref{sd}.  For our results, 
$|r| \approx 0.7$ gives a very good agreement, which is, on the other 
hand, theoretically very unlikely.  Recently, an admixture of $s$-wave 
component with $r = -0.067$ for $\kappa$-(BEDT-TTF)Cu({NCS})$_2$ was 
suggested\cite{Won:PHB:SC} from the analysis of the angular dependent 
magnetothermal conductivity data\cite{Izawa:PRL:NCS}.  These data 
suggest that the $d_{x^2-y^2}$-wave instead of $d_{xy}$-wave pairing 
is responsible for superconductivity in the $\kappa$-(ET)$_2 X$ 
materials.  This implies that both Fermi surfaces participate in 
pairing in contrast to previous assumption\cite{Louati:SYM:ET}.  Since 
in the former case the spin fluctuation model for the $\kappa$-(ET)$_2 X$ 
materials no longer describes the 
superconductivity pairing, the origin should be found elsewhere.  
Coulomb interaction, which is responsible for the $d$-wave 
superconductivity, gives rise to both spin and charge fluctuations, 
so the obvious solution appears to be
that charge fluctuations play the principal role in the
$\kappa$-(ET)$_2 X$ superconductivity.  The value $r = -0.07$ suggests
that the nodal lines in $\Delta(\vec{k})$ pass through the gap
between two Fermi surfaces.  This is consistent with the
$d+s$-wave model in which the superconductivity is due to 
charge fluctuations between different groups of dimers.
On the other hand, for $r \approx -0.7$,
the nodal directions cross the 2D circular Fermi surface,
and for $r \approx 0.7$, the nodal directions cross the 1D 
Fermi surface.  If $d+s$ superconductivity model is generated by 
charge fluctuations, such a scenario is unlikely to work, since 
this implies the strong intra Coulomb repulsion in each energy band.

Now, we address the behavior of the in-plane superfluid density 
in the intermediate state (Fig.~\ref{rSchR}).
Considering the polynomial fit in Eq.~(\ref{rsin3}), we see that the
in-plane superfluid density appears to fit very well the $d$-wave 
model with impurity scattering in the unitary
limit\cite{Hirschfeld:PRB:dw,Sun:PRB:dw,Sun:EPL:dw}.  However, there is a 
serious discrepancy between our experimental data and the impurity 
model.  First of all, we ascribe the difference between A~state 
(ground state), R~state (intermediate state) and Q~state to the
residual degree of the ethylene disorder.  Annealed state we discussed is 
the ground state with the least disorder in the ethylene groups. 
It is natural to assume that the ethylene disorder gives
rise to the quasiparticle scattering, which changes the $t$-linear
dependence of $\rho_{\text{s,in}}$ into the $t$-squared dependence.
But this is in contradiction with the well established fact that a
small disorder depresses $T_{\text{C}}$ 
dramatically\cite{Sun:PRB:dw,Sun:EPL:dw}.  In our experiment, it seems 
that $T_{\text{C}}$ is practically unaffected by the ethylene disorder 
for the intermediate state, achieved by slow cooling of -0.2~K/min.\ in 
the region of the glass transition.  Further, taking into account the 
fact that the $\lambda(0)$ increases for at least by a factor of 6, 
superconducting electron density at 0~K, 
$n_{\text{s}}(0) \propto \lambda^{-2}(0)$, decreases at least
by a factor of 36.  A simple impurity model cannot describe the
surprising combination of these two features.  On the other hand, 
the results for other low-temperature states achieved by cooling rates 
$q_{\text{C}} < -1 $~K/min.\ for sample~S1, and for Q~state in 
sample~S2 are more consistent with the theory.  In these cases both 
$T_{\text{C}}$, $\chi'(0)$ 
and therefore $\lambda(0)$ are concomitantly influenced by the 
remnant ethylene disorder.  However, it is still difficult to 
correlate the observed behavior to the impurity model quantitatively.
This discrepancy might indicate that the degree of disorder at low
temperatures, as defined by the cooling rate in the region of the glass
transition, has also a profound influence on electronic correlations,
responsible for SC pairing.  At this point we would like to recall the
result of Kund \textit{et al.}\cite{Kund:PHB:Br} showing changes in the
crystal structure parameters in the region of the glass transition.
It might be that these changes are also susceptible to the cooling rate.
In addition, authors of Ref.~\onlinecite{Taniguchi:PRB:Br} have pointed
out that the role of disorder 
in this class of superconductors might be different than in the
other unconventional superconductors due to the vicinity of the
Mott insulator in the phase diagram.

Coming back to the intermediate state, we suggest that the 
indication of the decrease in the superconducting electron 
density may be related to the reduction of the 
superconductivity volume.  It has been already reported that
cooling rates combined by progressive deuteration influence the 
low-temperature electronic state in $\kappa$-(ET)$_2$Br 
samples\cite{Kawamoto:PRB:Br,Taniguchi:PRB:Br}.  Deuterated 
$\kappa$-(ET)$_2$Br system is situated in the critical region between 
an insulating AF transition at 15~K and a SC transition at 11.5~K.  
Despite the slow cooling rate, the deuterated sample gives almost the same 
$T_{\text{C}}$ as in the hydrogenated $\kappa$-(ET)$_2$Br system, the SC 
state is not fully established in the bulk.  Note that this 
results strongly resembles what we observed for S2 samples in 
the intermediate (R) state.  In addition, a gradual decrease of 
the susceptibility below $T_{\text{C}}$ in the deuterated system 
strongly indicates the inhomogenous nature of the SC state.  This 
is in contrast to what we observe in sample~S2, in which the 
susceptibility curves are rather sharp even in Q~state. More 
rapid cooling rates induce a decrease of $T_{\text{C}}$ and a 
substantial decrease of SC volume fraction. Authors of 
Refs.~\onlinecite{Kawamoto:PRB:Br,Taniguchi:PRB:Br} have argued 
that since the electronic specific heat of rapidly cooled 
deuterated samples did not show any finite electronic 
contribution at low temperatures, the missing part of 
superconducting phase should be considered to be the magnetic 
insulating phase. The question arises if their conclusion might also 
be valid in the hydrogenated system.  Taking into account the 
structure of the phase diagram of this class of 
superconductors\cite{Lefebvre:PRL:Cl}, we think that this might 
be the case.  However, specific heat data 
under carefully controlled cooling cycles are needed to resolve this 
issue.

The most intriguing fact about Fig.~\ref{rSchR} is that in 
the intermediate state the observed data could be well described by the 
$s$-wave model as well.  This gives a possible explanation for the
contradictory findings in favor of the $s$-wave and $d$-wave model in
the same material.  The behavior is obviously strongly influenced
by both thermal history as well as synthesis, which 
suggests that the same material was not measured in the same
low-temperature state.  We hope that our results could 
contribute to reconcile contradictory findings met in the past.

\section{Conclusion}

The level of residual disorder and electronic properties at low 
temperatures are critically determined by the time scale of experiment 
in the region of the glassy transition and the sample synthesis.  This
fact imposes an additional requirement to get a reliable description 
of the SC state and that is to perform a full characterization of the SC 
state in the sample under study in the same well defined and controlled 
cooling conditions.  The origin of the observed differences can be
attributed to the residual ethylene disorder, which might
be theoretically considered as the impurity effect.

The in-plane superfluid density of the ground state with the
lowest residual ethylene disorder exhibits
clear $T$-linear dependence, which is consistent with the $d$-wave 
model and in contradiction with the $s$-wave model.   The
leading $T$-linear coefficient is much larger than the one
expected for the weak-coupling $d$-wave model.  The d+$s$-wave
model can remove this numerical discrepancy.  However,
rather large $s$-wave component needed to fit our data is not
consistent with the recent thermal conductivity results.

On the other hand, the in-plane superfluid density of the
intermediate state achieved by slow cooling of -0.2~K/min.\ 
in samples of one synthesis, as well as in the states achieved 
by rapid cooling rates ($q_{\text{C}} < -1 $~K/min.) in all 
studied samples, clearly exhibits the $T$-squared dependence, 
consistent with $d$-wave superconductivity in presence of 
impurities.  In the intermediate state, the same 
$T_{\text{C}}$ as in the ground state is accompanied by a 
relatively large reduction in the superconducting electron 
density, which cannot be explained within the standard 
impurity model.  In the rapidly cooled states,
$T_{\text{C}}$ does reduce concomitantly with $\lambda(0)$, 
still the depression of $T_{\text{C}}$ is quantitatively too small 
and the reduction of the superconducting electron density 
is too large.  More work, both theoretical and experimental, 
is needed to be done to resolve these issues.

Finally, the in-plane superfluid density data in the intermediate 
state can be relatively well fitted to the $s$-wave model data as 
well.  This fact gives a possible explanation for the contradictory 
findings in favor of $s$-wave and $d$-wave model in the same 
material and will hopefully contribute to a long awaiting consensus 
regarding the pairing symmetry in the $\kappa$-(BEDT-TTF) based 
superconductors.  In this circumstance, specific heat measurements
in the same well defined and controlled cooling conditions are highly
desirable.

\begin{acknowledgments}
The authors are grateful to K. Kanoda and D. Schweitzer for supplying
single crystals of different syntheses used in this research
and for very useful discussions.  We also thank O. Milat and 
V. Bermanec for photographing the samples.
\end{acknowledgments}

\end{document}